\documentclass[prd,twocolumn,superscriptaddress,showpacs,nofootinbib,
preprintnumbers]{revtex4}

\usepackage{amsmath}
\usepackage{amsfonts}
\usepackage{graphicx}
\usepackage{dcolumn}

\def\be{\begin{equation}}
\def\ee{\end{equation}}
\def\ba{\begin{eqnarray}}
\def\ea{\end{eqnarray}}
\def\bs{\begin{subequations}}
\def\es{\end{subequations}}

\usepackage{color}


\begin{document}

\title{A primer on problems and prospects of dark energy }
\author{M. Sami}
\affiliation{Centre for Theoretical Physics, Jamia Millia Islamia,
New Delhi-110025, India}
\begin{abstract}
This review on dark energy is intended for a wider audience,
beginners as well as experts. It contains important notes on various
aspects of dark energy and its alternatives. The section on
Newtonian cosmology followed  by heuristic arguments to capture the
pressure effects allows us to discuss the basic features of physics
of cosmic acceleration without actually resorting to the framework
of general theory of relativity. The brief discussion on
observational aspects of dark energy is followed by a detailed
exposition of underlying features of scalar field dynamic relevant
to cosmology. The review includes pedagogical presentation of
generic features of models of dark energy and its possible
alternatives.
\end{abstract}

\maketitle
\section{Introduction}
Twentieth century has witnessed remarkable developments in the field
of cosmology. The observation of redshift of light emitted by
distant objects and the discovery of microwave background in 1965
have revolutionized our thinking about universe. The hot big bang
model then received the status of standard model of universe.
However, in spite of the theoretical and observational successes,
cosmology remained confined to a rather narrow class of scientist;
others considered it as the part of a respectable philosophy of
science. Cosmology witnessed the first revolution in 1980 with the
invent of cosmological inflation making it acceptable to the larger
community of physicists. Since then it goes hand in hand with high
energy physics. The scenario envisages that universe has gone
through a phase of fast accelerated expansion at early epochs.
Inflation is a beautiful paradigm which can resolve some of the in
built inconsistencies of the hot big bang model and provides a
mechanism for generation of primordial fluctuations needed to seed
the structure we see in the universe today. In the past two decades,
observations have repeatedly confirmed the predictions of inflation.
However, its implementation is {\it ad hoc} and requires support
from a fundamental theory of high energy physics. As inflation takes
place around the Planck epoch, the needle of hope points towards
string theory $-$ a {\it consistent theory of quantum gravity}.

The second revolution cosmology witnessed in 1998, is related to
late time cosmic acceleration\cite{PR1,PR2}. The observations of
high redshift supernovae reveals that universe is accelerating at
present. The phenomenon is indirectly supported by data of
complimentary nature such as CMB, large scale structure, baryon
acoustic oscillation and weak lensing. It is really interesting that
the thermal history of our universe is sandwiched between two phases
of accelerated expansion.
 In the Newtonian language, cosmic repulsion can be realized by
supplementing the Newtonian force by a repulsive term on
phenomenological grounds. The rigorous justification of the
phenomenon can only be provided in the frame work of general theory
of relativity (see Ref.\cite{earlyde} for early attempts in this
direction). Late time acceleration can be fueled either by an exotic
fluid with large negative pressure dubbed {\it dark
energy}\cite{review1,vpaddy,review2,review3,review3C,review3d,review4}
or by modifying the gravity itself\cite{FRB}. The simplest candidate
of dark energy is provided by cosmological constant
$\Lambda$\cite{zelvarun,rev0,rev0B}, though, there are difficult
theoretical issues associate with it. Its small numerical value
leads to {\it fine tuning} problem and we do not understand why it
becomes important today {\it a la} {\it coincidence} problem.

Scalar fields provide an interesting alternative to cosmological
constant\cite{Paul,Kes}. To this effect, cosmological dynamics of a
variety of scalar fields has been investigated in the literature(see
review\cite{review2} for details). They can mimic cosmological
constant like behavior at late times and can provide a viable
cosmological dynamics at early epochs. Scalar field models with
generic features are capable of alleviating the fine tuning and
coincidence problems. As for the observation, at present, it is
absolutely consistent with $\Lambda$ but at the same time, a large
number of scalar field models are also permitted. Future data should
allow to narrow down the class of permissible models of dark energy.

As an alternative to dark energy, the large scale modifications of
gravity could account for the current acceleration of universe. We
know that gravity is modified at short distance and there is no
guarantee that it would not suffer any correction at large scales
where it is never verified directly. Large scale modifications might
arise from extra dimensional effects or can be inspired by
fundamental theories. They can also be motivated by phenomenological
considerations such as $f(R)$ theories of gravity. However, any
large scale modification of gravity should reconcile with local
physics constraints and should have potential of being distinguished
from cosmological constant. To the best of our knowledge, all the
schemes of large scale modification, at present, are plagued with
one or the other problems.

The review is organized as follows: After introduction and a brief
background, we present cosmology in Newtonian framework in section
III and mention efforts to put it on the rigorous foundations in the
domain of its validity. In section IV, we put forward heuristic
arguments to incorporate $\Lambda$, in particular and pressure
corrections, in general, in the evolution equations and describe the
broad features of cosmological dynamics in presence of cosmological
constant. In section V, we provide a short introduction to
relativistic cosmology and discus issues associated with
cosmological constant. After a brief subsection on observational
aspects of cosmic acceleration, we proceed to highlight the generic
features of scalar field dynamics relevant to cosmology and mention
the current observational status of dynamics of dark energy. In the
last section before summary, we present a discussion on the current
problems of alternatives to dark energy.

Last but not the least, a suggestion for the follow up of this
review is in order. At present, there exist, a number of excellent
reviews on dark
energy\cite{review1,vpaddy,review2,review3,review3C,review3d,review4}
and cosmological constant\cite{zelvarun,rev0,rev0B} which focus on
different aspects of the subject. Four recent and very interesting
reviews\cite{review3,review3C,review3d,review4} which try to address
the theoretical and observational aspects of late time cosmic
acceleration are highly recommended. Humility does not allow to say
that Ref.\cite{review2} is the most comprehensive theoretical review
on dark energy with pedagogical exposition.

\section{The smooth expanding universe}
Universe is clumpy at small scales and consists of very rich
structure of galaxies, local group of galaxies, clusters of
galaxies, super clusters and voids. These structures typically range
from kiloparsecs to 100 megaparsecs. The study of large scale
structures in the universe shows no evidence of new structures at
scales larger than 100 megaparsecs. Universe appears smooth at such
scales which leads to the conclusion that universe is homogeneous
and isotropic at large scales which serves as one of the fundamental
assumptions in cosmology known as {\it cosmological
principal}\cite{Bondi}. Homogeneity tells us that universe looks the
same observed from any point whereas isotropy means that lt looks
same in any direction. In general these are two independent
requirements. However, isotropy at each point is stronger assumption
which implies homogeneity also. Cosmological principal presents an
idealized picture of universe which allows us to understand the
background evolution. The departure from smoothness can be taken
into account through perturbations around the smooth background.
Observations confirm the presence of tiny fluctuations from
smoothness in the early universe. According to modern cosmology,
these small perturbations via gravitational instability are believed
to have grown into the structures we see today in the
universe\cite{Weinberg,Padmanabhan,
Narlikar,Zel,TC,Vol3,LL,dodelson,mukhanov}.

One of the most remarkable discoveries in cosmology includes the
expansion of universe and its beginning from the big bang. The
analysis of radiation spectrum emitted from distant galaxies shows
that wavelengths of spectral lines are larger than the actually
emitted ones, the phenomenon is known as redshift of light. The
redshift is quantified by symbol $z$ defined as,
$z=(\lambda_{ob}-\lambda_{em})/\lambda_{em}$. According to the
Doppler effect, the wavelength of light emitted by a source receding
from the observer appears shifted towards red end of spectrum and
the redshift is related to the velocity of recession $v$ as, $z
\simeq v/c$ for $v<<c$. In the beginning of the last century,
astronomers could measure the distances to a number of distant
galaxies. Hubble carried out investigations of recession velocities
and plotted them against the distances to galaxies. He concluded in
1929 that there is a linear relation between recession velocity of
the galaxies and the distance to them $-$ the so called {\it Hubble
Law}.

The observational conclusion that universe expands is based upon the
redshift of radiation emitted by distant galaxies. Can we have
another explanation for the redshift? It might look surprising that
photons from larger distances emitted from galaxies reach us
redshifted due to the recession of galaxies and nothing else happens
to them. They travel through intergalactic medium and could be
absorbed by matter present there and then emitted loosing part of
their energy in this process thereby leading to their redshift
without resorting to expansion of universe. This apprehension can be
refuted by a simple argument. As for the absorption, the underlying
process is related to the scattering of photons by the particles of
intergalactic medium. If true, the source should have appeared
blurred which is never observed. Other efforts assuming the exotic
interactions of photons could not account for the observed redshift.
Thus the only viable explanation of the phenomenon is provided by
the expansion of universe\cite{Zel}.

If we imagine moving backward in time, universe was smaller in size,
the temperature was higher and there was an epoch when the universe
was vanishingly small with infinitely large energy density and
temperature-the beginning of universe dubbed as the {\it big bang}.
The matter was thrown away with tremendous velocity, since then the
universe is expanding and cooling. At early times it was extremely
hot and consisted of a hot plasma of elementary particles, there
were no atoms and no nuclei. Roughly speaking, at temperatures
higher than the binding energy of hydrogen atom, the photons were
freely scattering on electrons and atoms could not form. As the
universe cools below the temperature characterized by the binding
energy of hydrogen atom, electrons combine with protons to form
hydrogen atom leading to the decoupling of radiation from matter.
This was an important epoch in the history of universe known as
recombination. The decoupled radiation since than is just expanding
with the expanding universe and cooling. The discovery of microwave
background, the relic of the big bang, in 1965 confirms the
hypothesis of hot big bang.
\section{The homogeneous and isotropic Newtonian cosmology}
 Newtonian theory of gravitation allows us to understand the
expansion of a homogeneous isotropic universe in a simple way.
Newtonian description is valid provided the matter filling the
universe is non-relativistic and scales associated with the problem
are much smaller than the Hubble radius. For instance, at early
epochs, the universe was hot dominated by radiation. Hence early
universe strictly speaking, should be treated by relativistic
theory. The general theory effects are also crucial at super Hubble
scales. Despite its limitations, Newtonian cosmology provides a
simple and elegant way of understanding the expansion of
universe\cite{Bondi,Zel,mukhanov,Cahill}.
\subsection{Hubble law as a consequence of homogeneity and isotropy}
Using the Newtonian notions of physics, let us show that the Hubble
law is a natural consequence of homogeneity and isotropy. Let us
choose a coordinate system with origin $O$ such that matter is at
rest there and let us observe the motion of matter around us from
this coordinate system. The velocity field, i.e., the velocity of
matter at each point $p$ around us at an arbitrary time, depends
upon the radius vector {\bf r} and time $t$. We should now look for
the most general velocity field in a homogeneous and isotropic
universe. Let us assume another observer located at point $O'$ with
radius vector ${\bf r}_{o'}$ and moving with velocity ${\bf v}({\bf
r}_{o'})$ with respect to the observer $O$. If we denote the
velocity of point $p$ relative to $O$ and $O'$ at time $t$ by ${\bf
v}({\bf r}_p)$ and ${\bf v'}({\bf r}_p')$, we have,

\begin{eqnarray}
&&{\bf r}_p'={\bf r}_p-{\bf r}_{o'} \\
&& {\bf v'}({\bf r}_p')={\bf v}({\bf r}_p)-{\bf v}({\bf r}_{o'})
\end{eqnarray}
 where ${\bf r}_p$ and ${\bf r}_p'$ denote the radius vectors of point $p$
 with respect to $O$ and $O'$ respectively. The cosmological principal tells that the velocity field should have
the same functional form at any point,
\begin{eqnarray}
  {\bf v}({\bf r}_p')={\bf v}({\bf r}_p)-{\bf v}({\bf r}_{o'})
\end{eqnarray}
which clearly implies that the velocity field is a linear function
of its argument ${\bf r}$,
\begin{equation}
{\bf v}({\bf r},t)=T(t) {\bf r}
\end{equation}
where $T$ is a $3\times 3$ matrix. The matrix can always be
diagonalized by choosing a suitable coordinate system. Isotropy then
reduces it to kronecker symbol ($T_{i,j}=H(t) \delta_{i,j}$) leading
to
\begin{equation}
{\bf v}({\bf r,t})=H(t) {\bf r} \label{HubbleN1}
\end{equation}
where $H$ is known as the Hubble parameter.
 In general, a velocity field can
always be decomposed into rotational part, inhomogeneous part and
isotropic part at each point. It is not then surprising that the
homogeneous and isotropic velocity field has the form
(\ref{HubbleN1}) known as Hubble law.

It can easily be verified that Hubble law holds at any point. If we
move from $O$ to $O'$, we can write
\begin{equation}
{\bf v'}({\bf r}_p')= H{\bf r}_p-H{\bf r}_{o'}=H(t) {\bf r}_p'
\label{Hlaw}
\end{equation}
The Hubble's law gives the most general form of velocity field
permissible by the homogeneity and isotropy of space.

Hubble law tells us how distance between any two points in space
changes with time provided we know the expansion rate given by
$H(t)$,
\begin{equation}
 {\bf r}(t)={\bf x}e^{\int_0^t{H(t) dt}},~{\bf x} \equiv {\bf
r}(t=0)
 \label{Eqr}
\end{equation}
The law of expansion depends upon how the Hubble parameter $H$
varies with time. Eq. (\ref{Eqr}) shows how distances in a
homogeneous and isotropic universe scale with the scale factor
$a(t)$,
\begin{eqnarray}
&&a(t)\equiv e^{\int_0^t{H(t) dt}}~~or~~H(t)=\frac{\dot{a}}{a} \\
&&r(t)=a(t)x \label{pcdis}
\end{eqnarray}
The complete information of dynamics of a homogeneous and isotropic
universe is contained in the scale factor; we, thus, need evolution
equation to determine $a(t)$. In case $H$ is independent of time we
have exponentially expanding universe dubbed de-Sitter space. In
what follows we shall confirm that constant Hubble rate is allowed
in relativistic cosmology provided the energy density of matter in
the universe is constant. It is believed that universe has passed
through an exponentially expanding phase known as {\it inflation} at
early times.

According to Hubble law, in a homogenous and isotropic universe, all
the material particles move away radially from the observer located
at any point in the universe. This motion is refereed to as Hubble
flow. Indeed, any freely moving particle in such a background would
ultimately follow the Hubble flow. Motion over and above the Hubble
flow is called peculiar motion which can only arise in a perturbed
universe. It often proves convenient to change a coordinate system
dubbed {\bf comoving} which expands with expanding universe. Matter
which follows the Hubble flow will be at rest in the comoving
coordinate system, i.e. matter filling a homogeneous isotropic
universe is at rest with respect to the comoving observer. Both the
frames are physically equivalent. Let us clarify that universe does
not appear homogeneous and isotropic to any observer; for instance
if an observer is moving with a large velocity say towards a
particular galaxy, universe looks different to him/her. A physical
coordinate system is the system in which matter is at rest at the
origin and moves away radially at other points. The radius vector
${\bf r}$ of any point in this system called physical, changes with
time whereas its counterpart ${\bf x}$ in the comoving system is
constant. This means that physical distance between any two points
in the expanding universe is given by the comoving distance
multiplied by a factor which depends upon time which is precisely
expressed by Eq.(\ref{Eqr}) or equivalently by Eq.(\ref{pcdis}).
\subsection{Evolution equations}
We now turn to the evolution equation for the scale factor. Thank to
isotropy, we can employ spherical symmetry to derive the evolution
equation. At a given time $t$ called the cosmic time, let us
consider a sphere centered at $O$ with radius $r(t)$. Let
$\rho_b(t)$ be the  density of matter in the homogeneous isotropic
space referred to as background space hereafter. We assume that the
net gravitational force on a particle of mass $m$ situated on the
surface of the sphere due to matter out side the sphere is zero
which means that matter inside the sphere alone can influence the
motion of the particle. The total energy of the particle on the
surface of the sphere (see Fig.{\ref{Sphere}) at any time is
constant given by the expression\cite{Milne},


\begin{figure}
\resizebox{5cm}{!}{\includegraphics{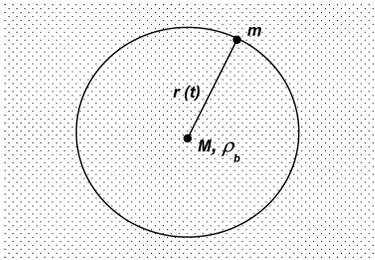}} \caption{Particle of
mass $m$ on the surface of a sphere of radius $r(t)$ in an expanding
universe with uniform matter density.} \label{Sphere}
\end{figure}
\begin{equation}
E_{Tot}=\frac{1}{2}m\dot{r}^2-\frac{4\pi}{3}mG\rho_br^2
\label{EtotN1}
\end{equation}
This equation can be cast in the following convenient form,
\begin{equation}
H^2\equiv \left(\frac{\dot{r}(t)}{r(t)}\right)^2=\frac{8 \pi}{3}G
\rho_b(t)+\frac{2E_{Tot}}{m r^2(t)} \label{FriedmannN1}
\end{equation}
which readily translates into an evolution equation for $a(t)$ (see,
Eq.(\ref{pcdis})) known as Friedmann equation,

\begin{equation}
H^2\equiv\left(\frac{\dot{a}}{a}\right)^2=\frac{8 \pi}{3}G
\rho_b(t)-\frac{K }{a^2},~~~~~~~~K=-\frac{2E_{Tot}}{x^2m}
\label{FriedmannN2}
\end{equation}
where $K$ can be zero, negative or positive depending how kinetic
energy compares with the
 potential energy.

In order to solve the evolution equation for $a(t)$, we need to know
how matter density $\rho_b(t)$ changes with time, i.e., we need the
conservation equation in the expanding universe. For
non-relativistic fluid, the continuity equation that gives us the
evolution of matter density of the fluid is,
\begin{equation}
\frac{\partial \rho_b(t)}{\partial t}+ \left(\nabla.\rho_b{\bf
v}\right)=0 \label{EqcontN0}
\end{equation}
Remembering that the matter density of the background fluid is
independent of the coordinates and the fluid velocity is given by
the Hubble law (\ref{Hlaw}), we transform the continuity equation to
have the usual form,
\begin{equation}
\frac{\partial \rho_b(t)}{\partial t}+ 3 H\rho_b=0 \label{EqcontN01}
\end{equation}
which formally integrates to,
\begin{equation}
\rho_b(t)=\rho_b^{(0)} \left(\frac{a_0}{a}\right)^3
\label{densityevol}
\end{equation}
where the subscript '0' denotes the quantities at the present epoch.
The evolution of matter density of non-relativistic fluid has a
simple meaning that mass of fluid in a co-moving volume remains
constant.

 Though the Eq.(\ref{FriedmannN2}) formally resembles the
 evolution equation of relativistic cosmology, its {\it derivation
 presented above is defective}. The expression for the potential
 energy is written with an assumption that gravitational potential
 can be chosen zero at infinity which is not true in an infinite
 universe. Since the mass density $\rho_b$ is constant in space, the total mass of universe
 diverges as $r^3$. As a result, the potential $-4\pi G \rho_b r^2/3$
 can not be normalized to zero at $r=\infty$.
 One could try to circumvent the problem by assuming
 that $\rho_b$ vanishes for a given large value of $r$ but it would conflict
 with the underlying assumption of homogeneity. Therefore, {\it conservation of energy
 is difficult to understand in an infinite universe with uniform
 matter density}.

We can also derive the evolution equations using the Newtonian force
law\cite{Bondi}. The force on the unite mass situated on the surface
of homogeneous sphere with radius $r$ is given by
\begin{equation}
{\bf F}=-\frac{4 \pi G}{3}\rho {\bf r} \label{FN1}
\end{equation}
The Euler's equation
\begin{equation}
\frac{\partial{\bf v}}{\partial t}+({\bf v}.{\bf \nabla}){\bf
v}=-\frac{{\bf \nabla}P_b}{\rho_b}+{\bf F}
\end{equation}
in a homogeneous isotropic background simplifies to
\begin{equation}
{\bf F}=(\dot{H}+H^2){\bf r} \label{hdot}
\end{equation}
where ${\bf F}$ is the force per unit mass on the fluid element
given by Eq.(\ref{FN1}). We have used the fact that pressure
gradients are absent in a homogeneous isotropic background and the
velocity field is given by the Hubble law. It should also be noted
that the pressure $P_b=0$ for the non-relativistic background fluid
under consideration. Using expressions (\ref{FN1}) $\&$
(\ref{hdot}), we obtain the equation for acceleration,
\begin{equation}
\frac{1}{a}\frac{d^2 a}{dt^2}=-\frac{4\pi G}{3} \rho_b(t)
\label{EqacN1}
\end{equation}
which could also be obtained directly from Eq.(\ref{FN1}). Equation
(\ref{EqacN1}) can easily be integrated to give the Friedmann
equation. Indeed, by multiplying the above equation by $\dot{a}$ and
using the evolution of mass density allows us to write
\begin{eqnarray}
\label{EqHuubleN1}
&&H^2=\frac{8\pi G}{3}\rho_b(t)-\frac{K}{a^2} \\
&&K \equiv a_0^2 \left(\frac{8\pi G \rho_0}{3}-H_0^2\right)
\label{KN}
\end{eqnarray}
The above {\it derivation is also problematic} as it assumes that
mass out side the sphere, used while writing Eq.(\ref{FN1}), can be
neglected which is not true for infinite universe with constant mass
density.

The problem can be circumvented by using the geometric reformulation
of Newtonian gravity in the language of Cartan. {\it According to
Cartan's formulation, orbits of particles are assumed to be the
geodesics of an affine space and gravity is then described by the
curvature of the affine connection}(see Ref.\cite{Tipler} and
references therein). According to Ref.\cite{Tipler}, no pathology in
cosmology associated with Newton's force law then occurs and the
evolution equations of {\bf Newtonian cosmology},
\begin{eqnarray}
\label{EqHubbleN11}
&&H^2=\frac{8\pi G}{3}\rho_b(t)-\frac{K}{a^2}\\
&& K \equiv a_0^2 \left(\frac{8\pi G \rho_b^0}{3}-H_0^2\right)\\
 && \frac{1}{a}\frac{d^2 a}{dt^2}=-\frac{4\pi G}{3} \rho_b(t) \label{EqacN11} \\
&& \frac{\partial \rho_b(t)}{\partial t}+ 3 H\rho_b(t)=0
\label{EqcontN011}
\end{eqnarray}
can be put on rigorous foundations. Eqs.(\ref{EqHubbleN11}),
(\ref{EqacN11}), $\&$ (\ref{EqcontN011}) are identical to evolution
equations of Friedmann cosmology for non-relativistic fluid filling
the universe. Whether or not one adopts the formulation presented in
Ref.\cite{Tipler}, Newtonian cosmology is nevertheless elegant and
simple.

Let us point out an important feature of Newtonian cosmology. We
note that the expression of $K/a^2$ remains unchanged under the
scale transformation $a(t) \to C a(t)$, $C$ being constant. As a
result, the evolution equations (\ref{EqacN11}) $\&$
(\ref{EqHubbleN11}) also respect the scale invariance. This
invariance is a characteristic of specially flat Friedmann
cosmology. The Newtonian cosmology can mimic all three topologies of
relativistic cosmology corresponding to $K=0,\pm 1$ in spite of the
fact that the underlying geometry in Newtonian cosmology is
Euclidean.
 Let us note that
the scale factor in Newtonian cosmology can always be normalized to
a convenient value at the present epoch. This is related to a simple
fact that the Friedmann equation (\ref{EqHubbleN11}) does not change
if we re-scale the scale factor which leaves the normalization of
$a$ arbitrary. The often used normalization fixes the scale factor
$a(t)=1$ at the present epoch, i.e, $a_0=1$. In case of relativistic
cosmology, the latter can only be done in case of $K=0$ whereas in
case of $K=\pm 1$, the numerical value of the scale factor $a_0$
depends upon the matter content of universe.

The second important feature of Newtonian cosmology is that it leads
to an evolving universe. Indeed, we could ask for a static solution
given by $\dot{a}(t)$ and $\ddot{a}(t)=0$ which is permitted by the
Friedmann equation (\ref{EqHubbleN11}) but not allowed by the
equation for acceleration (\ref{EqacN11}). It is really remarkable
that {\it Newtonian cosmology gives rise to an evolving universe}.
It is an irony that the discovery of expansion of universe had to
wait the general theory of relativity. This is related to the
commonly held perceptions of static universe which was prevalent
before Friedmann discovered the non-static cosmological solution of
Einstein equations. So much so that Einstein himself did not believe
in the Friedmann solution in the beginning and tried to reconcile
his theory with static universe by introducing cosmological constant
which he later withdrew.

\subsection{The past, the future and how old are we?}
 The general features of solutions of evolution equations can be
 understood without actually solving them. What can we say about the past and the
 fate of universe? The equation for acceleration tells us that $\ddot{a}<0$ for standard form of matter.
 This means that $a(t)$ as a function of time is concave downward.
 We need input regarding $\dot{a}$ at present to make important
 conclusion about the past. Observation tells us that $\dot{a}(t)>0$ at
 present. Thus $a(t)$ monotonously decreases as $t$ runs backward.
 It is therefore clear that there was an epoch in the history of
 universe when $a(t)$ vanishes identically. Without the loss of
 generality we can take $t=0$ corresponding to $a(t)=0$.

 As for the fate of universe, the problem is
 similar to that of escape velocity, namely, if $K>0$, the kinetic
  energy is less than the potential energy. In this case $a(t)$
 would increase to a maximum value where $\dot{a}(t)=0$, it would start
 decreasing thereafter till it vanishes and universe ends itself in {\it big crunch}. In case, $K<0$, scale
 factor would go on increasing for ever; $K=0$ represents the
 critical case. Three different possibilities, $K=0$, $K >0$ or
 $K<0$
 correspond to {\it critical}, {\it closed} and {\it open} universe
 respectively. We should emphasize that the fate of universe
 also crucially depends upon the nature of matter filling the
 universe. In some case, the universe may end itself in
 a singular state or the {\it cosmic doomsday}.

 Which of the three possibilities is realized in nature? To answer this question, let us rewrite
 Eq.(\ref{FriedmannN2}) in a convenient form,
 \begin{eqnarray}
\Omega_b(t)-1=\frac{K}{(aH)^2},~
\Omega_b(t)=\frac{\rho_b(t)}{\rho_c(t)}
 \label{OmegaH}
 \end{eqnarray}
 where the critical density is defined as, $\rho_c(t)=3H^2(t)/8\pi G$. Specializing the expression
 (\ref{OmegaH}) to the present epoch, we find that,
\begin{eqnarray}
&&\Omega_b^{(0)}>1~ (\rho_b^{(0)}>\rho_c^{(0)}) \Rightarrow K>0 \to~ closed~universe,\nonumber\\
&&\Omega_b^{(0)}=1~ (\rho_b^{(0)}=\rho_c^{(0)}) \Rightarrow K=0 \to~critical~universe, \nonumber\\
&&\Omega_b^{(0)}<1~ (\rho_b^{(0)}<\rho_c^{(0)})\Rightarrow K<0
\to~open~universe. \nonumber
\end{eqnarray}}
where the super script $'0'$ designates the corresponding physical
quantities at the present epoch. Since we know the observed value of
$\rho_c^{(0)}$, one of the three types of universe we live in,
depends upon how matter density in universe compares with
$\rho_c^{(0)}$. Observations on Cosmic Microwave background (CMB)
indicate that universe is critical to a good accuracy or $K \simeq
0$ which is consistent with inflationary paradigm.

Let us come to the solution of Newtonian cosmology in case of $K=0$.
Substituting $\rho_b(t)$ from Eq.(\ref{densityevol}) in
Eq.(\ref{EqHubbleN11}), we find that, ${\dot{a}^2}\sim a^{-1}$ which
easily integrates giving rise to
\begin{eqnarray}
&&a(t) =\left(\frac{t}{t_0}\right)^{2/3} \\
&&\rho_b(t)=\rho_b^{(0)}\left(\frac{t_0}{t}\right)^2 \\
&&H(t)=\frac{2}{3}\frac{1}{t} \label{HubbleN5}
\end{eqnarray}
The above solution is known as Einstein-de-Sitter solution. We can
estimate the age of universe using Eq.(\ref{HubbleN5}),
\begin{equation}
t_0=\frac{2}{3}\frac{1}{H_0} \label{ageN}
\end{equation}
Interestingly, if gravity were absent, universe would expand with
constant rate given by $H_0$. Using Hubble law we would then find,
\begin{equation}
t_0=\frac{1}{H_0}
\end{equation}
which is the maximum limit for the age of universe in the hot big
bang model ($2H_0^{-1}/3\leq t_0<H^{-1}$). The presence of standard
matter always leads to deceleration thereby leading to smaller time
taken to reach the present Hubble rate of expansion. The presence of
cosmological constant or any other exotic form of matter can
crucially alter this conclusion.

\subsection{Cosmological constant {\it a la} Hooke's law}
We have seen that Newtonian cosmology gives rise to evolving
universe but for the historical reasons, cosmology had to wait the
general theory of relativity to discover it. The fact that Newtonian
cosmology leads to non-stationary solution was known before general
theory was discovered but it could receive attention as it
conflicted with perception of static universe. Attempts were then
made to modify Newtonian gravity to reconcile it with the static
universe. Clearly, the modification should be such that it becomes
effective at large scales leaving local physics unchanged. Looking
at the Newton's force law (\ref{FN1}), it is not difficult to guess
that static solution is possible provided that we add a repulsive
part proportional to the radius vector ${\bf r}$ in Eq.(\ref{FN1}).
Newton's law of gravitation should therefore be supplemented by
linear force law\cite{Neumann,St,Bondi,Zel}
\begin{equation}
{\bf F}=-\frac{4 \pi G}{3}\rho_b {\bf r} +\frac{1}{3} \Lambda {\bf
r}\label{FN11}
\end{equation}
where $\Lambda$ is known as cosmological constant which is positive
in the present context. It is interesting to note that there are
only two central forces namely, the inverse square force and the
linear force which give rise to stable circular orbits.

Our discussion of cosmological constant is heuristic and the cheap
motivation here is to incorporate the repulsive effect in the
evolution equations. We rewrite the modified force law (\ref{FN11})
as an equation of acceleration using the comoving coordinates,
\begin{eqnarray}
 && \frac{1}{a}\frac{d^2 a}{dt^2}=-\frac{4\pi G}{3} \rho_b(t) +\frac{\Lambda}{3} \label{EqacN12}
\end{eqnarray}
which shows that a positive $\Lambda$ term contributes to
acceleration as it should. The integrated form of Eq.(\ref{EqacN12})
is given by,
\begin{eqnarray}
\label{EqHubbleN12} &&H^2=\frac{8\pi
G}{3}\rho_b(t)-\frac{K}{a^2}+\frac{\Lambda}{3}
\end{eqnarray}
where the integration constant $K$ can be formally written again
through physical quantities defined at the present epoch. The
modified force law (\ref{FN11}) was proposed much before Einstein's
general theory of relativity by Neumann and Seeliger in
1895-96\cite{Neumann,St}.

 Let us note that adding cosmological constant to Newtonian
force is equivalent to adding a {\it constant matter density}
$\rho_{\Lambda}=\Lambda/8\pi G$ to the background matter density
$\rho_b$ which does not to go well with the continuity equation
(\ref{EqcontN01}). Since the acceleration equation also gets
modified in presence of $\Lambda$, we should check whether the
modified evolution equations allow this possibility. If we
differentiate Eq.(\ref{EqHubbleN12}) with respect to time and
respect the modified acceleration equation, we find that constant
matter density is permissible in the expanding universe. As for the
continuity Eq.(\ref{EqcontN01}), it is valid for a perfect
non-relativistic fluid. The cosmological constant does not belong to
this category, the pressure corresponding to constant energy density
is not zero. The continuity equation should take the note of
pressure and get appropriately modified. As pointed out earlier our
present discussion of cosmological constant here is qualitative.
Rigorously speaking, we are trying to get the right thing in the
wrong place! We shall come back to this point after we incorporate
the pressure corrections in the evolution equations.

Evolution equations (\ref{EqHubbleN12}) and (\ref{EqacN12}) admit a
static solution ($a=const=a_0$) in case of $K>0$. Static Einstein
universe ( $\dot{a}=0$ and $\ddot{a}=0$) is possible provided that
$\Lambda$ has definite numerical value
\begin{eqnarray}
\Lambda=\Lambda_{c}={4\pi G}\rho_b^{(0)}
\end{eqnarray}
We shall observe after a short while that the static Einstein
universe is unstable under small fluctuations.

The qualitative features of solutions of evolution equations can be
understood without actually solving them. Eq.(\ref{EqacN12}) can be
thought as an equation of a point particle in one
dimension\cite{review1,vpot},
\begin{equation}
\ddot{a}=-\frac{\partial V}{\partial a} \label{gradpot}
\end{equation}
moving in potential field
\begin{equation}
V(a)=-\left(\frac{4\pi G\rho_b a^2}{3}+\frac{\Lambda a^2}{6}\right),
\end{equation}
where we have used the fact that $\rho_b \sim a^{-3}$. The Hubble
equation acquires the form of the total energy of the mechanical
particle
\begin{equation}
E=\frac{\dot {a}^2}{2}+V(a) \label{EqEN1}
\end{equation}
where $E=-K/2$. In order to make the mechanical analogy transparent,
let us compute the minimum of the kinetic energy. If the minimum
exists, it should obviously correspond to the numerical value of the
scale factor that gives rise to the maximum of the effective
potential $V(a)$. It is easy to see that the kinetic energy is
minimum if $a=a_m$,
\begin{eqnarray}
&&a_m=(A/\Lambda)^{1/3}  \\
 && \left(\frac{\dot {a}^2}{2}\right)_m=\frac{1}{2}\left(A^{2/3}
\Lambda^{1/3}-K\right), \label{KeminN1}
\end{eqnarray}
where $A=4\pi G \rho_b^{(0)}a_0^3$. Note that $V(a)$ is maximum at
$a=a_m$. From Eq.(\ref{KeminN1}), we infer that the kinetic energy
of the system at the top of the potential is,
\begin{eqnarray}
\left(\frac{\dot {a}^2}{2}\right)_m \geq 0~if~ \Lambda \geq
\Lambda_c\equiv \frac{K^3}{A^2}
\end{eqnarray}
In case $\Lambda=\Lambda_c$, the system barely makes to the hump of
the potential ($\dot{a}=0)$ corresponding to $a_m=a_0$ where
$\ddot{a}=0$ as it should be (see Eq.(\ref{gradpot}) which is
nothing but the Einstein's static solution. We are now ready to
provide the qualitative description of solutions of evolution
equations. For $\Lambda<\Lambda_c$, the kinetic energy is formally
negative for $a=a_m$ which means that it vanishes before the
particle reaches the maximum of the potential.
\begin{figure}
\resizebox{8cm}{!}{\includegraphics{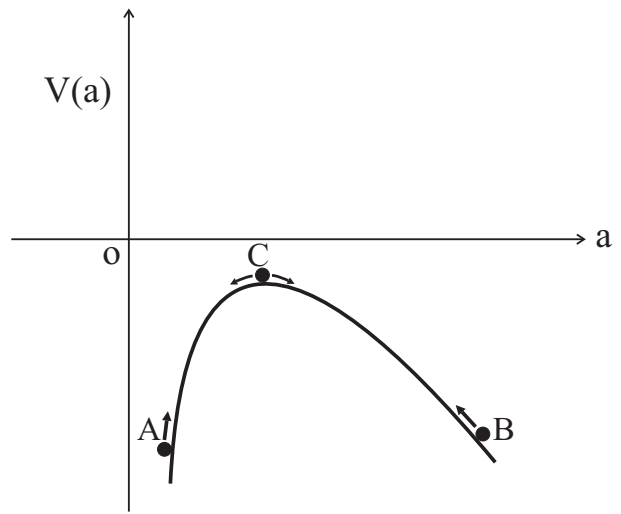}} \caption{Plot of the
effective potential $V(a)$ versus the scale factor $a$.
Configurations (A) $\&$ (B) correspond to motion of system beginning
from $a=0$ and $a=\infty$ respectively. (C) corresponds to static
solution unstable under small fluctuations. } \label{figplot}
\end{figure}
 In Fig.\ref{figplot} we have
displayed the plot of $V(a)$ versus the scale factor $a$. We show
three possible configuration of interest: (A) Corresponds to motion
starting from the left of the barrier with $a=0$. (B) Depicts the
situation in which the potential barrier is approached beginning
from the right with a large value of the scale factor. (C)
Represents the possibility of static
solution. \\

We first analyse the case of $K>0$ or $E<0$ which gives rise
 to a variety of interesting possibilities.\\
1. $\Lambda< \Lambda_c$: In this case, the kinetic energy is
insufficient to overcome the potential barrier giving rise to the
following
interesting solutions.  \\
 $\bullet$ (a)
Oscillating solution: In this case, motion starts from $a=0$ with
insufficient kinetic energy to reach the hump of the potential. In
this situation, the scale factor increases up to a maximum  value
where $\dot{a}=0$ for $a <a_m$ marking the turning point followed by
the
contraction to $a=0$.\\
$\bullet$ (b) Bouncing universe:  If the potential barrier is
approached from the right side with $a=\infty$, the scale factor
first decreases and reaches a minimum value and then bounces to
expanding phase as the kinetic energy is not enough to
overcome the barrier.\\
 $\bullet$ (c) Einstein static universe : This configuration
corresponds to the maximum of the potential with  $\dot{a}=0$ and
$\ddot{a}=0$, possible for a particular value of $\Lambda$, obtained
earlier. Clearly, static universe corresponding to point particle
sitting on the hump of the potential, is not stable. Small
perturbations would derive it to either contracting ($a\to 0)$ or
expanding $(a\to
\infty)$ universe.\\

2. {\bf $\Lambda> \Lambda_c$}: The kinetic energy is sufficient to
overcome the barrier for this choice of $\Lambda$. As a result,
motion first decelerates till the system reaches the top of the
potential and then slides down the hill with acceleration. Scale
factor exhibits the point of inflection at $a(t)=a_m<a_0$. If
$\Lambda$ slightly exceeds its critical value, an interesting
possibility dubbed {\it loitering universe} can be realized. The
scale first increases as it should, approaches $a_0$ and remains
nearly frozen for a substantial period before entering the phase of
acceleration. Such a scenario has important implications for
structure formation.

For $K\leq 0$ or $E\geq 0$, the system always has enough kinetic
energy to surmount the barrier allowing the scale factor to increase
from $a=0$ to large values as time increases. This case is similar
to the one with $K>0$ and $\Lambda>\Lambda_c$.

For any given value of $\Lambda$, the scale factor exhibits the
point of inflection at $a=a_m=(4\pi G\rho^{(0)}_b
a_0^3/\Lambda)^{1/3}$. This also clear from Eq.(\ref{FN11}) $\&$
(\ref{EqacN12}) in which the first term is of attractive character
and dominates in the beginning leading to deceleration. However, as
the scale factor increases and reaches a particular value, the
repulsive term takes over; the scale factor exhibits the point of
inflection and the expansion becomes accelerating thereafter.

Observations should tell us when deceleration changed into
acceleration. This crucially depends upon how $4\pi G\rho^{(0)}_b$
compares with $\Lambda$ or how $\rho_{\Lambda}$ compares with
$\rho_M^{(0)}/2$. The transition from deceleration to acceleration
should have taken place around the present epoch. Had it happened
much earlier it would have obstructed structure formation\cite{WS}.
We shall come back to this point to confirm that cosmic acceleration
is indeed a recent phenomenon.
\section{Beyond Newtonian physics: Pressure corrections} The
formalism of Newtonian cosmology is not applicable to relativistic
fluids. Relativistic fluids essentially have non-zero pressure. For
instance, radiation is a relativistic fluid with pressure
$P_b=\rho_b c^2/3$. The cosmological constant also belongs to the
category of relativistic systems. In general theory of relativity,
pressure appears on the same footing as energy density. Here we
present heuristic arguments to capture the pressure corrections in
the evolution equations (see Ref.\cite{Zel}).

Let us consider a unit comoving volume in the expanding universe and
assume the expansion to be adiabatic. The first law of
thermodynamics then tells that
\begin{equation}
dE+P_b dV=0 \label{EN1}
\end{equation}
where $P_b(t)$ is the  pressure of background fluid. The first law
of thermodynamics applies to any system, be it relativistic or
non-relativistic, classical or quantum $-$ {\it thermodynamics is a
great science}. \\
The energy density of the fluid can
always be expressed through the mass density, \begin{equation}
E=\frac{4\pi}{3}a^3\rho_bc^2 \label{EN2}
\end{equation}
Substituting (\ref{EN2}) into (\ref{EN1}), we obtain the continuity
equation in the expanding universe,
\begin{equation}
\dot{\rho_b}+3H\left(\rho_b+\frac{P_b}{c^2}\right)=0
\label{ConseqNR1}
\end{equation}
Thus the continuity equation responds to pressure corrections:
$\rho_b\to \rho_b+P_b/c^2$. For a non-relativistic fluid, rest
energy density dominates over pressure and the second term in the
parenthesis can be neglected. For instance, for dust, $P_b \simeq
0$. At early times, universe was hot and was dominated by radiation.
Hence the early universe should be treated by relativistic theory;
Newtonian description becomes valid at late times when matter
dominates. For the sake of convenience, we shall use the unit $c=1$.
With this choice, relativistic mass density and energy density are
same.

We can now present cosmological constant as a perfect fluid with
constant energy density. The continuity Eq.(\ref{ConseqNR1}) then
implies that $\rho_{\Lambda}=-P_{\Lambda}$. Next, we claim that the
correct equation of acceleration in case of background fluid with
energy density $\rho_b$ and pressure $P_b$ is given by
\begin{equation}
\frac{\ddot{a}}{a}=-\frac{4\pi
G}{3}\left(\rho_b+{3P_b}\right)+\frac{\Lambda}{3}
\label{AccelerationNR1}
\end{equation}
To verify, let us multiply Eq.(\ref{AccelerationNR1}) left right by
$\dot{a}$
\begin{equation}
\frac{1}{2}\frac{d}{dt}\left(\dot{a}^2\right)=-\frac{4\pi
G}{3}a\left(\rho_b\dot{a}+3P_b\dot{a}\right)+\frac{\Lambda}{3}a\dot{a}
\label{aAccelerationNR1}
\end{equation}
Using the continuity equation, we can express the term containing
pressure $P_b$ in Eq.(\ref{aAccelerationNR1}) through $\rho_b$,
$\dot{\rho_b}$ and $\dot{a}$
\begin{equation}
\frac{1}{2}\frac{d}{dt}\left(\dot{a}^2\right)=\frac{4\pi
G}{3}\frac{d}{dt}\left[\rho_b a^2+\frac{\Lambda}{6}a^2\right]
\label{Integral}
\end{equation}
which can put in form of Friedmann equation in the presence of
matter with non-zero pressure.
\begin{equation}
H^2=\frac{8\pi G}{3}\rho_b(t)-\frac{K}{a^2}+\frac{\Lambda}{3}
\label{HubbleRel}
\end{equation}
 We again observe that pressure corrects the energy
density. Positive pressure adds to deceleration where as the
negative pressure contributes towards acceleration, see
Eq.(\ref{AccelerationNR1}). It looks completely opposite to our
intuition that highly compressed substance explodes out with
tremendous impact whereas in our case pressure acts in the opposite
direction. It is important to understand that our day today
intuition with pressure is related to pressure force or pressure
gradient. In a homogeneous universe pressure gradients can not
exist. Pressure is a relativistic effect and can only be understood
within the frame work of general theory of relativity. Pressure
gradient might appear in Newtonian frame work in the inhomogeneous
universe but pressure can only be induced by relativistic effects.
Strictly speaking, it should not appear in Newtonian cosmology. This
applies to $\Lambda$ also with negative pressure which we introduced
in Newtonian cosmology by hand.
Eqs.(\ref{ConseqNR1}),(\ref{AccelerationNR1}) $\&$ (\ref{HubbleRel})
coincide with the evolution equations of relativistic cosmology.
Their derivation presented here is heuristic. The rigorous treatment
can only be given in the framework of general theory of relativity
where cosmological constant appears naturally.

In order solve the evolution equations, we need a relation between
the energy density and pressure known as equation of state. In case
of barotropic fluid the equation of state is given by
$w_b=P_b/\rho_b$. Dust and radiation correspond to $w_b=0,1/3$
respectively.
Assuming that universe is filled with perfect fluid with constant
equation of state parameter $w_b$, we find from
Eqs.(\ref{ConseqNR1}) $\&$ (\ref{EqHubbleN12}) in case of $K=0$,
\begin{eqnarray}
&& \rho_b \propto a^{-3(1+w)} \\
&& a(t) \propto t^{\frac{2}{3(1+w)}},~~(w>-1) \\
&& a(t) \propto e^{\sqrt{\frac{\Lambda}{3}}t}~~(w=-1)
\end{eqnarray}
In case of radiation, $w_b=1/3$ and as a result $\rho_b \equiv
\rho_r \propto {a^{-4}}$. In contrast to the case of dust dominated
universe, the radiation energy density decreases faster with the
expansion of universe. The positive radiation pressure adds to
energy density making the gravitational attraction stronger.
Consequently, the Hubble damping in the conservation equation
increases allowing the energy density decrease faster than dust in
expanding universe. This can also be understood in a slightly
different way, if we assume that radiation consists of photons. As
universe expands, the number density of photons scales as $a^{-3}$
as usual. But since any length scale in the expanding universe grows
proportional to the scale factor, the energy of a photon,
$hc/\lambda$ decreases as $1/a$ leading to $\rho_r\sim a^{-4}$ and
$a(t) \propto t^{1/2}$. It is clear that radiation dominated at
early epochs as $\rho_M \sim a^{-3}$ for dust.

Let us make an important remark on the dynamics in the early
universe which was dominated by radiation (for simplicity, we ignore
here  other relativistic degrees of freedom). As $\rho_r \sim
a^{-4}$, the first term on the RHS of evolution of Hubble equation
dominates over the curvature term $K/a^2$; obviously, cosmological
constant plays no role in the present case. We therefore conclude
that all the models effectively behave as $K=0$ model at early
times,
\begin{equation}
\frac{\dot{a}^2}{a^2}=\frac{8\pi G}{3}\rho_r^{(0)}\frac{a_0^4}{a^4}
\to \frac{a(t)}{a_0}=\left( \frac{32 \pi G
\rho_r^{(0)}}{3}\right)^{1/4} t^{1/2} \label{HuubleearlyN1}
\end{equation}
We next assume that radiation was in thermal equilibrium
characterized by the black body distribution,
\begin{equation}
\rho_r=b T^4, \label{StefN1}
\end{equation}
where $b$ is the radiation constant. From Eqs.(\ref{HuubleearlyN1})
$\&$ (\ref{StefN1}), we find how temperature scales with the
expansion of universe,
\begin{equation}
T=\left(\frac{\rho_r^0}{b}\right)^{1/4}\frac{a_0}{a}
\end{equation}
which on using Eq.(\ref{HuubleearlyN1}) tells us how early universe
cooled with time,
\begin{equation}
T= \left( \frac{32 \pi G }{3b}\right)^{1/4}  t^{-1/2}
\label{TemtimeN1}
\end{equation}
At $t=0$, both the radiation density and temperature become
infinitely large; all the physical quantities diverged at that time
referred to as {\it big bang}. The big bang singularity is not the
artifact of homogeneity and isotropy. It is a generic feature of any
cosmological model based upon classical general theory of
relativity. Classical physics breaks down
 as big bang is approached. In the framework of classical general relativity, the
 big bang is taken to be the beginning of our universe. Universe was
 thus born in a violent explosion like event throwing away cosmic
 matter and giving rise to expansion of universe. Since gravity is
 attractive (provided universe is filled with matter of non-negative pressure),
  its roll is to decelerate the expansion. What caused big bang, has no satisfactory
  answer. The big bang is a physical singularity which should be
  treated by quantum gravity. The inflationary paradigm can mimic
  big bang without singularity but in that case, we do not know what caused
  inflation!
In the cosmic history, there was an epoch when matter took over
leading to matter dominate era. It turns out that it took around
$10^{5}$ years for radiation energy density to equalize with energy
density of matter. The age of universe, i.e., the time elapsed since
the big bang till the present epoch given by (\ref{ageN}) changes
insignificantly, if we consider the universe filled with radiation
and dust both. This is because the time taken from the big bang till
radiation matter equality is negligibly small as compared to the
actual age of universe which is around 14 Gyr. Thus the age given by
(\ref{ageN}) is a reliable theoretical estimate. Unfortunately, the
age given by Eq.(\ref{ageN}) falls short than the age of some very
old objects found in the universe. This is one of the old problems
of hot big bang model. We shall discuss its possible remedy in the
dark energy dominated universe.
\subsection{Dark energy}
 Eqs.(\ref{HubbleRel}) and (\ref{aAccelerationNR1}) tell us that
 the positive cosmological constant $\Lambda$ contributes positively
 to the background energy density and negatively to pressure. It
 can be thought as a perfect barotropic fluid with,
 \begin{eqnarray}
\rho_{\Lambda}=\frac{\Lambda}{8\pi
G},~P_{\Lambda}=-\frac{\Lambda}{8\pi G}
 \end{eqnarray}
which corresponds to $w_{\Lambda}=-1$. In general, we find from
Eq.(\ref{AccelerationNR1}) that expansion has the character of
acceleration for large negative pressure,
\begin{eqnarray}
&&\frac{\ddot{a}}{a}=-\frac{4 \pi G}{3} \left(\rho_b +3P_b\right) \\
&& \ddot{a} > 0 \Rightarrow   P_b < -\frac{\rho_b}{3}:~ Dark~
energy.\nonumber
\end{eqnarray}
where we have included $\Lambda$ in the background fluid.
 Thus we need an exotic fluid dubbed {\it dark energy} to
fuel the accelerated expansion of universe. The various data sets of
complimentary support the late time acceleration of universe. The
simplest candidate of dark energy is provided by the cosmological
constant with $w_{\Lambda}=-1$ Observations at present do not rule
out the {\it phantom } dark energy with $w<-1$ corresponding to
super acceleration. In this case the expanding solution takes the
form,
\begin{eqnarray}
&& a(t)=(t_s-t)^n ,~~(n=2/3(1+w)<0) \\
&&H=\frac{n}{t_s-t}
\end{eqnarray}
where $t_s$ is an integration constant. It is easy to see that
phantom dominated universe will end itself in a singularity, in
future known as {\it big rip} or {\it cosmic doomsday} as $t\to
t_s$. Clearly, as $t\to t_s$, both the Hubble parameter and the
background energy density diverge\cite{CK}.
\subsection{Age crisis and its possible resolution}
Apart from the cosmic acceleration, dark energy has important
implications, in particular, in relation to the age problem. In any
cosmological model with normal form of matter, the age of universe
falls short compared to the age of some known objects in the
universe. Since the age of universe crucially depends upon the
expansion history, it can serve as an important check on the model
building in cosmology. In order to appreciate the problem, let us
first consider the case of flat dust dominated Universe
($\Omega_M=1$) in which case as shown earlier,
\begin{equation}
t_0=\frac{2}{3}\frac{1}{H_0}
\end{equation}
The observational uncertainty of $H_0$ gives rise to the following
estimate,
\begin{eqnarray}
&&H_0^{-1}=9.8 h^{-1} Gyr \\
&&0.64\lesssim h \lesssim 0.8 \to t_0=(8-10) Gyr
\end{eqnarray}
This model is certainly in trouble as its prediction for age of
Universe fails to meet the constrain following from the  study of
ages of old stars in globular clusters: $12 Gyr \lesssim t_0
\lesssim 15 Gyr$\cite{Kraus}. One could try to address the problem
by invoking the open model with $\Omega_M^{(0)}<1$. In this case the
age of universe is expected to be larger than the flat dust
dominated Universe $-$ {\it for less amount of matter, it would take
longer for gravitational attraction to slow down the expansion rate
to its present value}. Looking at Eq.(\ref{EqHubbleN11}), it is not
difficult to guess that in this case, $H_0 t_0 \to 1$ for
$\Omega_M^{(0)} \to 0$ which is a substantial improvement. However,
this model is not viable for several reasons. In particular, the
study of large scale structure and its dynamics constrain the matter
density: $0.2<\Omega_M^{(0)} <0.3$ and observations on CMB
un-isotropy reveal that universe is {\it critical } to a good
accuracy.

\begin{figure}
\includegraphics[width=6.0cm,angle=-90.0]{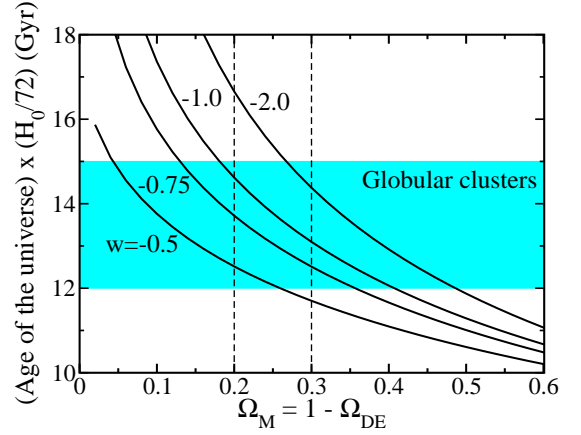}
\caption{Plot of age of Universe versus  $\Omega_M$ (at present
epoch) for a flat universe with matter and dark energy with constant
equation of state parameter $w$, from Ref.\cite{review4}}
\label{turnerfig}
\end{figure}

The age problem can be resolved in a flat universe dominated by {\it
dark energy}. Let us rewrite the Friedmann equation in a convenient
form,
\begin{equation}
\left(\frac{\dot{a}}{a}\right)^2=H_0^2
\left[\Omega_M^{(0)}\left(\frac{a_0}{a}\right)^3+\Omega_{DE}^{(0)}\left(\frac{a_0}{a}\right)^{3(1+w)}\right]
\label{EqHageN}
\end{equation}
which allows us to write the expression of $t_0$ in the closed form
\begin{equation}
t_0 =\frac{1}{H_0} \int_0^{\infty} {\frac
{dz}{(1+z)\left[\Omega_M^{(0)}(1+z)^3+\Omega_{DE}^{(0)}(1+z)^{3(1+w)}\right]^{1/2}}}
\label{ageintegN}
\end{equation}
where $\Omega_{M}^{(0)}$ is the contribution of dark matter and
$(1+z) \equiv a_0/a$, $z$ being the redshift parameter. The dominant
contribution to the age of universe comes from the matter dominated
era and we, therefore, have omitted $\Omega_r$ in
Eq.(\ref{EqHageN}).
 In case
dark energy is cosmological constant ($w_{\Lambda}=-1$), we get the
analytical expression for the age of Universe,
\begin{equation}
t_0=\frac{2}{3}\frac{H_0^{-1}}{\Omega_{\Lambda}^{1/2}}\ln\left(\frac{1+\Omega_{\Lambda}^{1/2}}
{{\Omega_M^{(0)}}^{1/2}}\right)
\end{equation}
For dark energy other than the cosmological constant, the integral
in Eq.(\ref{ageintegN}) should be computed numerically. In
Fig.\ref{turnerfig}, we have plotted the age of universe versus the
$\Omega_M^{(0)}$ for various possibilities of dark energy including
the phantom one. The age constraint can be met by flat dark energy
models provided that $-2~\lesssim w \lesssim -0.5$ for
$\Omega_M^{(0)}$ lying between $0.2$ and $0.3$, see
Ref.\cite{review4}. It is remarkable that hot big bang model can be
rescued by introducing the dark energy component. Interestingly,
cosmological constant was invoked to address the age problem before
the invention of cosmic acceleration. The observation of cosmic
acceleration in 1998 was a blessing in disguise for cosmological
constant.

\subsection{The discovery of cosmic acceleration and its confirmation}
\begin{figure}
\includegraphics[height=3.2in,width=3.2in]{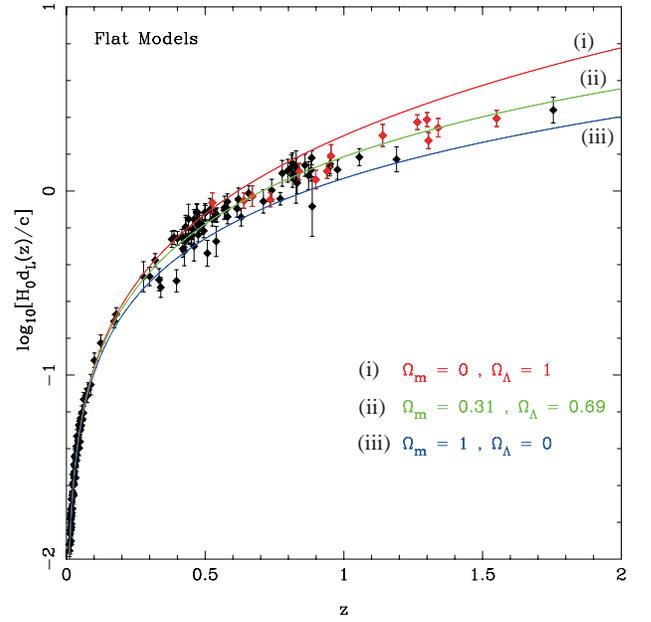}
\caption{Plot of the luminosity distance $H_{0}d_{L}$ versus the
redshift $z$ for a flat cosmological model. The black points come
from the ``Gold'' data sets by Riess {\it et al.} \cite{riessdata},
whereas the red points show the recent data from HST. Three curves
show the theoretical values of $H_{0}d_{L}$ for (i)
$\Omega_{M}^{(0)}=0$, $\Omega_{\Lambda}=1$, (ii)
$\Omega_{M}^{(0)}=0.31$, $\Omega_{\Lambda}=0.69$ and (iii)
$\Omega_{M}^{(0)}=1$, $\Omega_{\Lambda}=0$. {}From
Ref.~\cite{Cpaddy}.} \label{fitting}
\end{figure}

The direct evidence of current acceleration of universe is related
to the observation of luminosity distance by high redshift
supernovae by two groups independently in 1998. The luminosity
distance for critical universe dominated by non-relativistic fluid
and cosmological constant is given by
\begin{equation}
d_L=\frac{(1+z)}{H_0}\int_0^z{\frac{dz'}{\sqrt{\Omega_M^{(0)}\left(1+z'\right)^3+\Omega_{DE}^{(0)}(1+z')^{3(1+w)}}}}
\label{lumD}
\end{equation}
Eq.(\ref{lumD}) is the expanding universe generalization of absolute
 luminosity $L_s$ of a source and its flux ${\cal F}$ at a distance $d$  given by
 ${\cal F}=L_s/(4\pi d^2)$. It
follows from Eq.(\ref{lumD}) that $D_L \simeq z/H_0$ for small $z$
and that
\begin{eqnarray}
&&d_L=2\left(1+z-(1+z)^{1/2}\right)H_0^{-1},~~\Omega_{M}^{(0)}=1\\
&&d_L=z(1+z)H_0^{-1},~~~~\Omega_{DE}^{(0)}=\Omega_{\Lambda}=1
\end{eqnarray}
which means that luminosity distance at high redshift is larger in
universe dominated by cosmological constant which also holds true in
general for an arbitrary equation of state $w$ corresponding to dark
energy. Therefore supernovae would appear fainter in case the
universe is dominated by dark energy. The luminosity distance can be
used to estimate the apparent magnitude $m$ of the source given its
absolute magnitude $M$
\begin{equation}
m-M=5 \log\left(\frac{d_L}{Mpc}\right)+25 \label{magnitude}
\end{equation}
Let us consider two supernovae $1997{ap}$ at redshift $z=0.83$ with
$m=24.3$ and $1992 p$ at $z=0.026$ with $M=16.08$ respectively.
Since the supernovae are assumed to be the standard candles, they
have the same absolute magnitude. Eq.(\ref{magnitude}) then gives
the following estimate
\begin{equation}
H_0d_L\simeq 1.16
\end{equation}
Then theoretical estimate for the luminosity distance is given by
\begin{eqnarray}
&&d_L \simeq 0.95 H_0^{-1},~~\Omega_M^{(0)}=1\\
&&d_L \simeq 1.23 H_0^{-1},~~\Omega_M^{(0)}=0.3,
\Omega_{\Lambda}=0.7
\end{eqnarray}
 where we have used the fact that, $d_L \simeq z/H_0$ for small $z$.
 The above estimate lands a strong support to the hypothesis that late
 time universe is dominated by dark energy (see, Fig\ref{fitting}).

\begin{figure}
\includegraphics[width=6.0 cm, height=7.0cm]{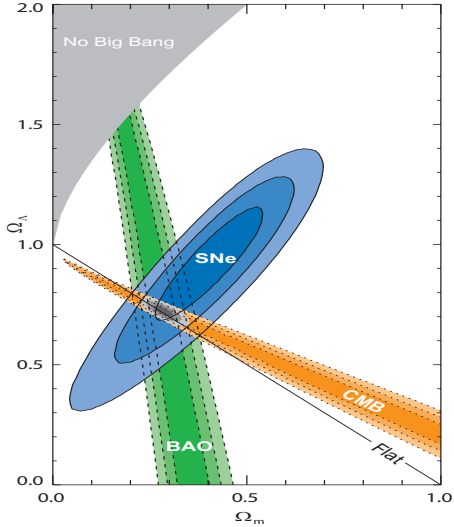}
\caption{Figure shows the best fit regions in the
($\Omega_{\Lambda},\Omega_M$) plane  obtained using the CMB, BAO and
supernovae data, from Ref.\cite{Kowalski}} \label{Kowalskifig1}
\end{figure}

\begin{figure}
\includegraphics[width=7.5cm,height=8.5cm]{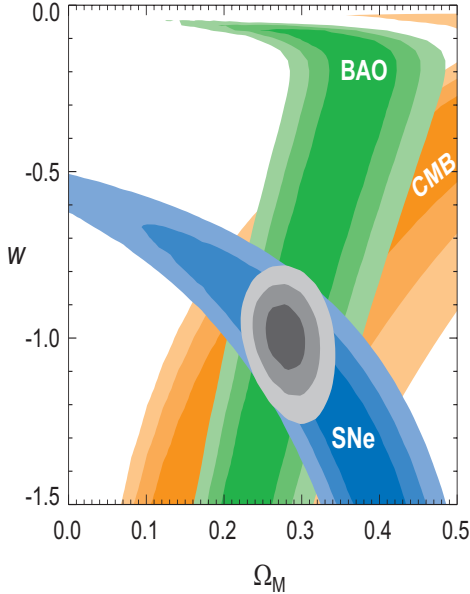}
\caption{Constraints on the dark energy equation of $w$ and
$\Omega_{M}$ obtained from CMB, BAO and supernovae observations,
from Ref.\cite{Kowalski}} \label{Kowalskifig2}
\end{figure}

 The observations related to CMB and large scale structure (LSS)
 provide an independent confirmation of dark energy scenario.
 The acoustic peaks of angular power spectrum of CMB temperature
 anisotropies contains important information. The location of the
 major peak tells us that universe is critical to a good accuracy
 which fixes for us the cosmic energy budget. Specializing the
 Friedmann Eq.(\ref{EqHageN}) to the present epoch ($a=a_0$), we have
\begin{equation}
\Omega_b^{(0)}=\Omega_M^{(0)}+\Omega_{DE}^{(0)}
\end{equation}
 The contribution of radiation to total fractional
 energy density $\Omega_b^{(0)}$ is negligible
 at present. The study of large
 scale structure and its evolution indicate that nearly $30\%$ of
 the total energy content is contributed by non-luminous component
 of non-barionic nature with dust like equation of state popularly
 known as {\it dark matter}. The missing component which is about
 $70\%$ is dark energy. The recent data on baryon acoustic
 oscillation is yet another independent probe of dark energy. The
 combined analysis of data of complimentary nature demonstrate that
 $\Omega_{DE}^{(0)}\simeq 0.7$ and $\Omega_{M}^{(0)}\simeq 0.3$, see Fig.\ref{Kowalskifig1}. The
 constraint on the equation of state parameter $w$ and $\Omega_{M}^{(0)}$
 shows that $w$ is restricted to a narrow strip around
 $w_{\Lambda}=-1$ (Fig.\ref{Kowalskifig2}). It is clear from the
 figure that the combined analysis allows super-negative values of
 $w$ corresponding to {\it phantom} energy.
Let us now confirm that the transition from deceleration to cosmic
acceleration took place in the recent past. Indeed, observations
allow to estimate the time of transition from deceleration to
acceleration. Let us rewrite Eq.(\ref{EqEN1}) through dimensionless
density parameters,
\begin{equation}
\frac{\dot{a}^2}{2}=\frac{H_0^2}{2}\left(\frac{\Omega_M^{(0)}a_0^3}{a}+\Omega_{\Lambda}a^2\right)
\label{EqEN2}
\end{equation}
Using Eq(\ref{EqEN2}), we can find out the numerical value of
$(a/a_0)$ corresponding to the minimum of kinetic energy
$(\dot{a}^2/2)$ which precisely gives the transition from
deceleration to acceleration,
\begin{equation}
\left(\frac{a}{a_0}\right)_{tr}=\left(\frac{\Omega_M^{(0)}}{2
\Omega_{\Lambda}}\right)^{1/3} \Rightarrow
z_{tr}=\left(\frac{2\Omega_{\Lambda}}{
\Omega_{M}^{(0)}}\right)^{1/3}-1\simeq 0.67
\end{equation}
for the observed values of density parameters ($\Omega_M^{(0)}\simeq
0.3$; $\Omega_{\Lambda}\simeq 0.7) $ and this confirms that the
contribution of $\Lambda$ to cosmic dynamics became important at
late times such that the cosmic acceleration is indeed a recent
phenomenon.
\section{Relativistic cosmology}
In the last section we presented heuristic arguments to capture the
pressure effects in the evolution equations. Pressure in cosmology
is a relativistic effect which can be consistently understood in the
frame work of general theory of relativity. Einstein equations are
complicated non-linear equations which do admit analytical solutions
in presence of symmetries. Homogeneity and isotropy of universe is
an example of a generic symmetry of space time. The assumption of
homogeneity and isotropy forces the metric to assume the FRW form
\begin{eqnarray}
&&ds^2=-dt^2+a^2(t)\left(\frac{dr^2}{1-Kr^2}+r^2(d\theta^2+\sin^2\theta d\phi^2)\right) \nonumber\\
&& K=0,\pm 1 \label{metric}
\end{eqnarray}
where $a(t)$ is scale factor. Coordinates (r,$\theta,\phi$) are the
comoving coordinates. A freely moving particle comes to rest in
these coordinates.

Eq.(\ref{metric}) is purely a kinemetic statement. The information
about dynamics is contained in the scale factor $a(t)$. Einstein
equations allow to determine the scale factor provided the matter
contents of universe is specified. Constant $K$ in the metric
(\ref{metric}) describes the geometry of the spatial section of
space time. $K=0,\pm 1$ corresponds to spatially flat, sphere like
and hyperbolic geometry respectively.

The differential equation for the scale factor follows from Einstein
equations
\begin{equation}
G_{\mu \nu}=R_{\mu \nu}-\frac{1}{2}g_{\mu\nu}R=8\pi G T_{\mu \nu}
\label{E1}
\end{equation}
where $G_{\mu \nu}$ is the Einstein tensor, $R_{\mu \nu}$ is the
Ricci tensor. The energy momentum tensor $T_{\mu \nu}$ takes a
simple form reminiscent of ideal perfect fluid in FRW cosmology
\begin{equation}
T_{\mu}^{\nu}=Diag(-\rho_b, P_b, P_b, P_b)
\end{equation}
Note that pressure in general theory of relativity appears on the
same footing as energy density. In the FRW background, the
components of $G_{\mu \nu}$ can easily be computed
\begin{equation}
G_0^0=-\frac{3}{a^2}\left(\dot{a}^2+K\right),
~~G_i^j=\frac{1}{a^2}\left(2a\ddot{a}+\dot{a}^2+K\right)
\end{equation}
Other components of $G_{\mu \nu}$ are identically zero. Einstein
equations then give rise to the following two independent equations
\begin{eqnarray}
\label{HR1}
&&H^2=\frac{8\pi G}{3}\rho_b-\frac{K}{a^2} \\
&& \frac{\ddot{a}}{a}=-\frac{4 \pi G}{3} \left(\rho_b+3 P_b\right)
\label{AcR}
\end{eqnarray}
We remind that $\rho_b$ designates the total energy density of all
the fluid components present in the universe. The continuity
equation $\dot{\rho_b}+3 H(\rho_b+P_b)=0$ can be obtained by using
Eqs.(\ref{HR1}) $\&$ (\ref{AcR}) which also follows naturally from
the Bianchi identity. As mentioned earlier, we can normalize the
scale factor to a convenient value at the present epoch in case of
specially flat geometry. In other cases, it should be determined
from the relation $a_0 H_0=\left(|\Omega_b^{0}-1|\right)$ where
$\Omega_b^{(0)}$ defines the total energy content of universe at the
present epoch.

Let us note that the Einstein equations (\ref{E1}) with the energy
momentum tensor of standard fluid with positive pressure can not
lead to accelerated expansion. The repulsive effect can be captured
either by supplementing the energy momentum tensor (on right hand
side of Einstein equations) with large negative pressure or by
modifying the geometry itself, i.,e. the left hand side of Einstein
equations. We can ask for a consistent modification of Einstein
equations (equation of motion should be of second order with the
highest derivative occurring linearly so that the Cauchy problem is
well posed) in four space time dimensions within the classical frame
work. Under the said conditions, the only admissible modification is
provided by the cosmological constant. Thus we can add a term
$\Lambda g_{\mu \nu}$ on the left hand side of Eq.(\ref{E1}) which
we can formally carry to the right hand side and interpret it as the
part of energy momentum tensor of a perfect fluid\cite{zelvarun}(see
also Refs.\cite{lothers} for a different approach to cosmological
constant),
\begin{equation}
G_{\mu \nu}=R_{\mu \nu}-\frac{1}{2}g_{\mu\nu}R=8\pi G T_{\mu
\nu}-\Lambda g_{\mu \nu}
\end{equation}
Such a modification is allowed by virtue of Bianchi identity. It is
remarkable that cosmological constant does not need {\it adhoc}
assumption for its introduction; it is always present in Einstein
equations. It could be considered as a fundamental constant of
classical general theory of relativity at par with Newton's constant
$G$. It is also interesting to note that model based upon
cosmological constant is consistent with all the observational
findings in cosmology at present. However, there are deep
theoretical problems related to cosmological constant.
\subsection{Theoretical issues associated with $\Lambda$}
There are important theoretical issues related to cosmological
 constant.
Cosmological constant can be associated with vacuum fluctuations in
the quantum field theoretic context\cite{zelvarun,rev0,rev0B}.
Though the arguments are still at the level of numerology but may
have far reaching consequences. Unlike the classical theory, the
cosmological constant in this scheme is no longer a free parameter
of the theory. Broadly the line of thinking takes the following
route.  The ground state energy dubbed zero point energy or vacuum
energy $\rho_{vac}$ of a free quantum field with spin $j$ given by
\begin{eqnarray}
&&\rho_{vac}=\frac{1}{2}\left(-1\right)^{2j}\left(2j+1\right)\int_0^{\infty}
{\frac{d^3{\bf k}}{2\pi^3}\sqrt{k^2+m^2}}\\
&&=\frac{\left(-1\right)^{2j}\left(2j+1\right)}{4\pi^2}\int_0^{\infty}{dk
k^2\sqrt{k^2+m^2}}
\end{eqnarray}
is ultraviolet divergent. This contribution is related the ordering
ambiguity of fields in the classical Lagrangian and disappears when
normal ordering is adopted. Since this procedure of throwing out the
vacuum energy is {\it adhoc}, one might try to cancel it by
introducing the counter terms. The later, however requires fine
tuning and may be regarded as unsatisfactory. The divergence is
related to the modes of very small wavelength. As we are ignorant of
physics around the Planck scale, we might be tempted to introduce a
cut off around the Planck length $L_{p}$ and associate with this a
fundamental scale. Thus we arrive at an estimate of vacuum energy
$\rho_{vac} \sim M_p^4$ (corresponding mass scale- $M_{vac} \sim
\rho_{vac}^{1/4}$) which is away by 120 orders of magnitudes from
the observed value of this quantity which is of the order of
$10^{-48}(GeV)^4$. The vacuum energy may not be felt in the
laboratory but plays important role in GR through its contribution
to the energy momentum tensor as
\begin{equation}
 < T_{\mu\nu} >_0= -
\rho_{vac} g_{\mu \nu},~ \rho_{vac}=\Lambda/8\pi G
\end{equation}
and appears on the right hand side of Einstein equations.

The problem of zero point energy is naturally resolved by invoking
supersymmetry which has many other remarkable features. In the
supersymmetric description, every bosonic degree of freedom has its
Fermi counter part which contributes zero point energy with opposite
sign compared to the bosonic degree of freedom thereby doing away
with the vacuum energy. It is in this sense the supersymmetric
theories do not admit a non-zero cosmological constant. However, we
know that we do not live in supersymmetric vacuum state and hence it
should be broken. For a viable supersymmetric scenario, for instance
if it is to be relevant to hierarchy problem, the suppersymmetry
breaking scale should be around $M_{susy}\simeq 10^3$ GeV. We are
still remain away from the observed value by many orders of
magnitudes. We do not know how Planck scale or SUSY breaking scales
is related to the observed vacuum scale!

 At present there is no satisfactory solution to
 cosmological constant problem. One might assume that there is some
 way to cancel the vacuum energy. One can then treat $\Lambda$ as a free
 parameter of classical gravity similar to Newton constant $G$.
 However, the small value of cosmological constant leads to several puzzles
 including the
  fine tuning and coincidence problems. The energy density in
  radiation at the Planck  scale is of the order of Plank energy density $\rho_P \simeq 10^{72} GeV^4$
  and the observed value of the dark energy density, $\rho_{\Lambda}\simeq 0.7 \times \rho_c^{(0)}\simeq 10^{-48} GeV^4$
  which implies that $\rho_{\Lambda}/\rho_P \sim
  10^{-120}$.
   Thus $\rho_{\Lambda}$ needs to be fine
  tuned at the level of one part in $10^{-120}$ around the Plank epoch,
  in order to match the current universe. Such an extreme fine
  tuning is absolutely unacceptable at theoretical grounds.
  Secondly, the energy density in cosmological constant is of the
  same order as matter energy density at the present epoch. The question what causes this
   {\it coincidence} has no satisfactory answer.

Efforts have recently been made to understand $\Lambda$ within the
frame work of string theory using flux compactification. String
theory predicts a very complicated landscape of  about $10^{500}$
de-Sitter vacua\cite{rev0B}. Using Anthropic principal, we are led
to believe that we live in one of these vacua!

A novel approach  to cosmological constant problem is provided in
Ref.\cite{paddylambda}. The line of thinking takes following route:
In the conventional framework, the equations of motion for matter
fields are invariant under the shift of the matter Lagrangian by a
constant while gravity breaks this symmetry. Thus, one cannot obtain
a satisfactory solution to the cosmological constant problem until
the gravity is made to respect the same symmetry. An effective
action suggested by Padmanbhan in Ref.\cite{paddylambda} is
explicitly invariant under the "shift symmetry". In his approach,
the observed value of the cosmological constant should arise from
the energy fluctuations of degrees of freedom located in the
boundary of a spacetime region.

\begin{figure}
\includegraphics[width=8.0cm,angle=00.0]{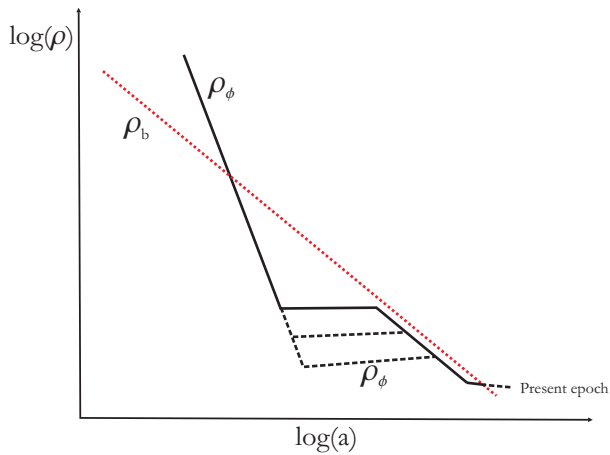}
 \caption{Cosmologically viable evolution of field energy density versus the scale factor. The dotted line shows
 the evolution of background (matter/radiation) energy density. The field energy
 density $\rho_{\phi}$ (with different initial conditions) joins the scaling regime
 and mimics the background. At late times it exits the scaling matter regime to become the dominant
 component and to account for the late time acceleration. }
\label{figshoot3}
\end{figure}

\begin{figure}
\includegraphics[width=8.0cm,angle=00.0]{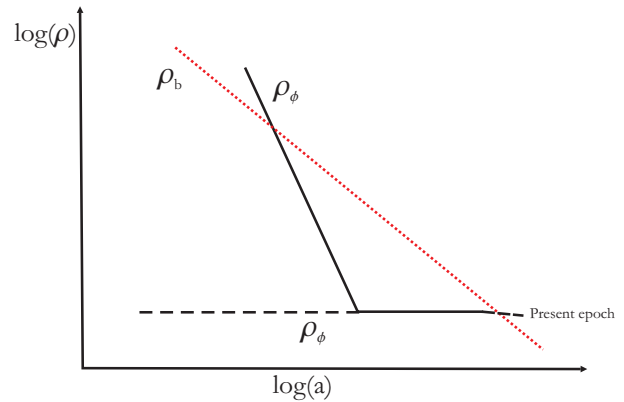}
 \caption{Evolution of $\rho_{\phi}$ and $\rho_b$ in absence of scaling regime
 in case of overshoot and undershoot. The field remains trapped in the locking
 regime till its energy density becomes comparable to that of the background component.
  It then starts evolving slowly and overtakes the background
  to become dominant at late times.}
\label{figshoot2}
\end{figure}
\section{Scalar field dynamics relevant to cosmology}
The fine tuning problem associated with cosmological constant led to
the investigation of cosmological dynamics of a variety of scalar
field systems such as quintessence, phantoms, tachyons and
K-essence\cite{Paul,Kes,scalar1,Lindercaldwell,scalar2}(see
review\cite{review2} for details). Scalar fields can easily mimic
dark energy at late times and posses rich dynamics in the past. We
should note that scalar fields models do not address the
cosmological constant problem, they rather provide an alternative
way to describe dark energy.
 The underlying dynamics of
these systems has been studied in great detail in the literature.
Scalar fields naturally arise in models of high energy physics and
string theory. It is worthwhile to bring out the broad features
their dynamics that
make these system viable to cosmology.\\
\subsection{Quintessence}
A standard scalar field (minimally coupled to gravity) capable of
accounting for the late time cosmic acceleration is termed as
quintessence. Its action is given by
\begin{equation}
S=\int{{\cal L}\sqrt{-g}d^4x}=-\int{\left(\frac{1}{2}g^{\mu
\nu}\partial_{\mu}\phi\partial_{\nu}
\phi+V(\phi)\right)\sqrt{-g}d^4x} \label{Qaction}
\end{equation}
The energy momentum tensor corresponding to this action is given by
\begin{equation}
T_{\mu \nu}=\partial_{\mu}\phi\partial_{\nu} \phi-g_{\mu
\nu}\left[\frac{1}{2}g^{\alpha
\beta}\partial_{\alpha}\phi\partial_{\beta} \phi+V(\phi)\right]
\end{equation}
which gives rise the following expression for energy density and
pressure in FRW background
\begin{equation}
\rho_{\phi}=\frac{1}{2}\dot{\phi}^2+V(\phi),~
P_{\phi}=\frac{1}{2}\dot{\phi}^2-V(\phi) \label{rhoPphi1}
\end{equation}
The Euler-Lagrangian equation
\begin{eqnarray}
&&\partial^{\alpha}\frac{ \delta\left(\sqrt{-g}{\cal L}\right)   }{
\delta\partial^{\alpha}\phi}-\frac{ \delta\left(\sqrt{-g}{\cal
L}\right)   }{ \delta\phi}=0  \\
&&\sqrt{-g}=a^3(t)
\end{eqnarray}
for the action (\ref{Qaction}) in FRW background acquires the form
\begin{equation}
\ddot{\phi}+3H\dot{\phi}+\frac {dV}{d\phi}=0
\end{equation}
which is formally equivalent to the continuity equation and can put
in the form
\begin{equation}
\rho_{\phi}=\rho_{\phi}^0
\exp{\left(-\int{3(1+w{(\phi)})\frac{da}{a}}\right)},
\label{rhoscal}
\end{equation}
where $w(\phi)=P_{\phi}/\rho_{\phi}$. Eq.(\ref{rhoPphi1}) tells us
that for a steep potential $\dot{\phi}^2>>V(\phi)$, the equation of
state parameter approaches the stiff matter limit, $w({\phi}) \to 1$
where as $w({\phi}) \to -1$ in case of a flat potential,
$\dot{\phi}^2<<V(\phi)$. Hence the energy density scales as
$\rho_{\phi} \sim a^{-n},~0\leq n\leq 6$. Let us note that while the
field rolls along the steep part of the potential, its energy
density $\rho_{\phi}$ scales faster than $\rho_r$.

 From Eq.(\ref{AcR}) we find that
\begin{equation}
\ddot{a} >0 \to \rho_b+3P_b <0 \Rightarrow \dot{\phi}^2<V(\phi)
\label{AcRC}
\end{equation}
which means that we need nearly flat potential to account for
accelerated expansion of universe such that
\begin{eqnarray}
\frac{1}{V}\left(\frac{V_{,\phi}}{V}\right)^2<<1,~~\frac{V_{,\phi\phi}}{V^2}<<1
\label{slowroll}
\end{eqnarray}
In case of field domination regime, the two conditions in
Eq.(\ref{slowroll}) define the slow roll parameters which allow to
neglect the $\ddot{\phi}$ term in equation of motion for $\phi$. In
the present context, unlike the case of inflation, the evolution of
field begins in the matter dominated regime and even today, the
contribution of matter is not negligible. The traditional slow roll
parameters can not be connected to the conditions on slope and
curvature of potential which essentially requires that Hubble
expansion is determined by the field energy density alone. Thus the
slow roll parameters are not that useful in case of late time
acceleration, though, Eq.(\ref{slowroll}) can still be helpful.

The scalar field model aiming to describe dark energy should possess
important properties allowing it to alleviate the {\it fine tuning}
and {\it coincidence} problems without interfering with the thermal
history of universe. The nucleosynthesis puts an stringent
constraint on any relativistic degree of freedom over and above that
of the standard model of particle physics. Thus a scalar field has
to satisfy several important constraints if it is to be relevant to
cosmology. Let us now spell out some of these features in detail,
see Ref.\cite{Paul,review2} for details. In case the scalar field
energy density $\rho_{\phi}$ dominates the background
(radiation/matter) energy $\rho_b$, the former should redshift
faster than the later allowing radiation domination to commence
which in tern requires a steep potential. In this case, the field
energy density overshoots the background and becomes subdominant to
it. This leads to the locking regime for the scalar field which
unlocks the moment the $\rho_{\phi}$ is comparable to $\rho_b$. The
further course of evolution crucially depends upon the form the
scalar potential. For the non-interference with thermal history, we
require that the scalar field remains unimportant during radiation
and matter dominated eras and emerges out from the hiding at late
times to account for late time acceleration. To address the issues
related to fine tuning, it is important to investigate the
cosmological scenarios in which the energy density of the scalar
field mimics the background energy density. The cosmological
solution which satisfy this condition are known as {\it scaling
solutions},
\begin{equation}
\frac{\rho_{\phi}}{\rho_b}=const.
\end{equation}
The steep exponential potential $V(\phi) \sim \exp(\lambda
\phi/M_P)$ with $\lambda^2>3(1+w_b)$ in the frame work of standard
GR gives rise to scaling solutions whereas the shallow exponential
potential with $\lambda \leq \sqrt{2}$ leads to a field dominated
solution ($\Omega_{\phi}=1$). Nucleosynthesis further constraints
$\lambda$. The introduction of a new relativistic degree of freedom
at a given temperature changes the Hubble rate which crucially
effects the neutron to proton ratio at temperature of the order of
one MeV when weak interactions freeze out. This results into a bound
on $\lambda$, namely\cite{review2},
\begin{eqnarray}
\Omega_{\phi}\equiv 3(1+w_b)/\lambda^2\lesssim 0.13 \Rightarrow
\lambda \gtrsim 4.5.
\end{eqnarray}
In this case, for generic initial conditions, the field ultimately
enters into the scaling regime, the attractor of the dynamics, and
this allows to alleviate the fine tuning problem to a considerable
extent. The same holds for the case of undershoot, see
Fig.\ref{figshoot3}.

Scaling solutions, however, are not accelerating as they mimic the
background (radiation/matter). One therefore needs some late time
feature in the potential. There are several ways of achieving this:
(1) The potential that mimics a steep exponential at early epochs
and reduces to power law type $V \sim \phi^{2 p}$ at late times
gives rise to accelerated expansion for $p<1/2$ as the average
equation of state $<w({\phi})>=(p-1)/(p+1)<-1/3$ in this
case\cite{wang,samioc1}. (ii) The steep inverse power law type of
potential which becomes shallow at large values of the field can
support late time acceleration and
 can mimic the background at early time\cite{samioc2}.

The solutions which exhibit the aforesaid features are referred to
as {\it tracker} solutions. For a viable cosmic evolution we need a
tracker like solution. However, on the basis of observations, we can
not rule out the non-tracker models  at present.

In the second class of models where trackers are absent, there are
two possibilities. First, if $\rho_{\phi}$ scales faster than
$\rho_b$ in the beginning, it then overshoot the background and
enter the locking regime. In case of the undershoot, the field is
frozen from the beginning due to large Hubble damping. In both the
cases, for a viable cosmic evolution, models parameters are chosen
such that $\rho_{\phi} \sim \rho_{\Lambda}$ during the locking
regime. Hence at early times, the field gets locked ($w{(\phi)}=-1$)
and waits for the matter energy density to become comparable to
field energy density which is made to happen at late times. The
field then begins to evolve towards larger values of $w{(\phi)}$
starting from $w{(\phi)}=-1$(see Fig.\ref{figshoot2}). In this case
one requires to tune the initial conditions of the field. The two
classes of scalar fields are called {\it freezing} and {\it thawing}
models\cite{Lindercaldwell,review3}. In case of tracker (freezing)
models, one needs to tune the slope of the field potential.
Nevertheless, these are superior to thawing models as they are
capable addressing both the fine tuning and the coincidence
problems.

Before we proceed further, we should make an honest remark about
scalar field models in general. {\it These models lack predictive
power: for a give cosmic history, it is always possible to construct
a field potential that would give rise to the desired evolution}.
Their merits should therefore be judged by the generic features
which arise in them. For instance tracker models deserve attention
for obvious reasons. Scalar fields inspired by a fundamental theory
such as {\it rolling tachyons} are certainly of interest.
\subsection{Tachyon field as source of dark energy}
Next we shall be interested in the cosmological dynamics of tachyon
field  which is specified by the Dirac-Born-Infeld (DBI) type of
action given by (see Ref.\cite{review2} and references therein),

\begin{equation}
\mathcal{S}=\int {
-V(\phi)\sqrt{1-\partial^\mu\phi\partial_\mu\phi}}\sqrt{-g} d^4x
\label{Taction1}
\end{equation}
where on phenomenological grounds, we shall consider a wider class
of potentials satisfying the restriction that $V(\phi) \to 0$ as
$\phi \to \infty$. In FRW background, the pressure and energy
density of $\phi$ are given by
\begin{equation}
P_{\phi}=-V(\phi)\sqrt{1-\dot{\phi}^2}
\end{equation}
\begin{equation}
\rho_{\phi}=\frac{V(\phi)}{\sqrt{1-\dot{\phi}^2}}
\end{equation}
The equation of motion which follows from (\ref{Taction1}) is
\begin{equation}
\ddot{\phi}+3H\dot{\phi}(1-\dot{\phi}^2)+\frac{V'}{V}(1-\dot{\phi}^2)=0
\end{equation}
where $H$ is the Hubble parameter
\begin{equation}
H^2=\frac{1}{3 M_p^2}\left(\rho_{\phi}+\rho_b\right)
\end{equation}
Tachyon dynamics is very different from that of the quintessence.
Irrespective of the form of its potential
\begin{equation}
w{(\phi)}=\dot{\phi}^2-1 \Rightarrow~-1\leq w{(\phi)} \leq 0
\end{equation}
The investigations of cosmological dynamics shows that in case of
tachyon field, there exists no solution which can mimic scaling
matter/radiation regime. These models necessarily belong to the
class of thawing models. Tachyon models do admit scaling solution in
presence of a hypothetical barotropic fluid with negative equation
of state. Tachyon fields can be classified by the asymptotic
behavior of their potentials for large values of the field: (i)
$V(\phi) \to 0$ faster then $1/\phi^2$ for $\phi \to \infty$. In
this case dark matter like solution is a late time attractor. Dark
energy may arise in this case as a transient phenomenon. (ii)
$V(\phi) \to 0$ slower then $1/\phi^2$ for $\phi \to \infty$ ; these
models give rise to dark energy as late time attractor. The two
classes are separated by $ V(\phi)\sim 1/\phi^2$ which is scaling
potential with $w{(\phi)}=const$.
These models suffer from the fine tuning problem; dynamics in this
case acquires dependence on initial conditions.
\subsection{Phantom field}
The scalar field models discussed above lead to $w{(\phi)}\geq -1$
and can not give rise to super acceleration corresponding to phantom
dark energy with $w{(\phi)} <-1$ permitted by observations, see
Fig\ref{Kowalskifig2}. The simplest possibility of getting phantom
energy is provided by a scalar field with negative kinetic energy.
Phantom field  is nothing but the Hoyle-Narlikar's creation field
(C-field) which was introduced in the steady state theory to
reconcile the model with the perfect cosmological principle. Though
the quantum theory of phantom fields is problematic, it is
nevertheless interesting to examine the cosmological consequences of
these fields at classical level. Phantom field is described by the
following action
\begin{equation}
S=\int{\left(\frac{1}{2}g^{\mu \nu}\partial_{\mu}\phi\partial_{\nu}
\phi-V(\phi)\right)\sqrt{-g}d^4x} \label{ActionPh}
\end{equation}
Its corresponding equation of state parameter is given by
\begin{equation}
w{(\phi)}=\frac{\frac{1}{2}\dot{\phi}^2+V(\phi)}
{\frac{1}{2}\dot{\phi}^2-V(\phi)} \label{wphiPh}
\end{equation}
which tells us the $w{(\phi)}<-1$ for $\dot{\phi}^2/2<V(\phi)$. An
unusual equation of motion for $\phi$ follows from (\ref{ActionPh})
\begin{equation}
\ddot{\phi}+3H\dot{\phi}-\frac {dV}{d\phi}=0 \label{fieldPh}
\end{equation}
It should be noted that evolution equation of phantom field is same
as that of the ordinary scalar field but with inverted potential
allowing the field with zero kinetic energy to rise up the hill. As
mentioned earlier, phantom energy is plagued with {\it big rip}
singularity which is characterized by divergence of the Hubble
parameter and curvature of space time after a finite interval of
time. In such a situation, quantum effects become important and one
should include higher curvature corrections to general theory of
relativity which can crucially modify the structure of the
singularity. To the best of our knowledge, the big rip singularity
can be fully resolved in the frame work of {\it loop quantum
cosmology}\cite{loop}. Big rip can also be avoided at the classical
level in a particular class of models in which potential has
maximum. In this case, the field  rises to the maximum of the
potential and ultimately settle on the top of the potential to give
rise to de-Sitter like behavior.

For a viable cosmic history, the phantom energy density similar to
the case of rolling tachyon should be subdominant at early epochs.
The field then remains frozen till late times before its energy
density becomes comparable to matter energy density. Its evolution
begins thereafter. Clearly, dark energy models based upon phantom
fields belong to the category of thawing models.
%
\subsection{Late time evolution of dark energy}
In the preceding subsections, we have described the cosmological
dynamics of quintessence, phantoms and rolling tachyon. These scalar
field models fall into two broad categories: (i) Tracker or freezing
models in which the field rolls fast at early stages such that it
mimics the background with $w_b=0$. At late times, $w(\phi)$ starts
deviating from dust like behavior and becomes negative moving
towards de-Sitter phase as the field rolls down its potential. (ii)
Non-tracker or thawing models are those in which the field is
trapped in the locking regime due to large Hubble damping such that
$w(\phi)=-1$. And only at late times, as $\rho_{\phi}$ becomes
comparable to the background energy density, the field begins to
evolve towards larger values of $w(\phi)$. As demonstrated by
Caldwell and Linder \cite{Lindercaldwell}, these models occupy
narrow regions in the $(w'\equiv dw/d\ln(a),w)$ plane,
\begin{eqnarray}
&&3w(1+w)<w'<0.2w(1+w)~~Freezing~models. \nonumber\\
&& 1+w<w'<3(1+w)~~~~~~~~~~~~Thawing~ models. \nonumber
\end{eqnarray}
where the upper and the lower bounds are obtained using analytical
arguments and numerical analysis of generic models belonging to both
the classes of models. As pointed out earlier (see
Fig.\ref{Kowalskifig2}), combined analysis of different observations
reveal that dark energy equation of state parameter lies in the
narrow strip around $w_{\Lambda}=-1$. The observational resolution
between the two classes of the model which is of the order of $1+w$
is therefore a challenge to future observations.

As mentioned earlier, the phantom and the tachyon dark energy models
belong to the class of thawing models. In this case, we can simplify
the dynamics around the present epoch by using the approximation
that $|1+w|<<1$ and that the slope of the potential is small. The
validity of the second approximation can be verified numerically in
each case. In this scheme of a plausible approximation, one arrives
at an amazing result: All the different dynamical systems, thawing
quintessence, phantom, tachyon and phantom tachyon follow a unique
evolutionary track. The distinction between the four classes of
scalar field systems and the distinction between different models
within each class is an effect of higher order than
$|1+w|$\cite{scalar2} which certainly throws a great challenge to
future generation experiments! Indeed, a recent examination of
observational data including 397 Type Ia supernovae at redshifts
$0.015 \leq z \leq 1.55$ has shown that evolving dark energy models
provide a slightly better fit to the data than the cosmological
constant \cite{shaf09}. If future data confirms this result then it
could mean that cosmic acceleration is currently slowing down which
may have important consequences for dark energy model building.
\subsection{ Quintessential inflation on brane: A beautiful model that does
not work}
\begin{figure}[htp]
\includegraphics[width=71mm]{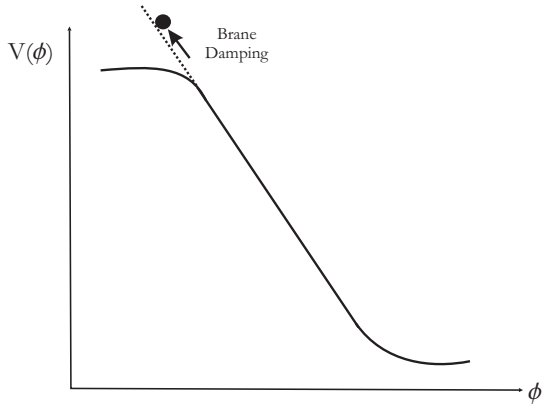}
\caption{ A desired form of potential for quintessential inflation.
It is generic to have a steep potential at early times with brane
corrections helping the slow-roll of the field.}\label{brane}
\end{figure}

 Quintessential inflation refers to attempts to describe
inflation and dark energy with a single scalar field. The
unifications of the two phases of accelerated expansion could be
realized in the framework of Randall-Sundrum (RS) brane
worlds\cite{samioc1,samioc2}. In order to achieve this, the field
potential should be flat during inflation but steep in radiation and
matter dominated eras such that $\rho_{\phi}$ could mimic the
background energy density at early epochs. At late times, it should
become flat so as to allow the current acceleration of universe.
Since the potential does not exhibit minimum, the conventional
reheating mechanism does not work in this scenario. One could employ
alternative mechanisms such as reheating via gravitational particle
production or instant preheating. It is not realistic to have  a
potential which changes from flat to steep and back to flat at late
times(see, Fig.\ref{brane}). However, it is generic to have a
potential which is steep and allows to track the background at early
epochs and gives rise to a viable late time cosmic evolution.

In case of a steep potential, the field energy density scales faster
that radiation energy density leading to the commencement of
radiative regime. But a steep potential can not support inflation in
FRW cosmology. This is precisely where the  brane assisted inflation
comes to our rescue. In $RS$ brane world model, the Friedmann
equation is modified to,
\begin{equation}
H^2=\frac{8\pi G}{3}\rho_b\left(1+\frac{\rho_b}{2\lambda_B}\right)
\end{equation}
where $\lambda_B$ is the brane tension. The presence of quadratic
density term in the Friedmann equation changes the dynamics at early
epochs in crucial manner. Consequently, the field experiences
greater damping and rolls down its potential slower than it would
during the conventional inflation. This effect is reflected in the
slow-roll parameters which have the form,
\begin{eqnarray}
&&\epsilon=\epsilon_{FRW}\frac{1+V/\lambda_B}{(1+V/2\lambda_B)^2}\\
&&\eta=\eta_{FRW}(1+V/2\lambda_B)^{-1}
\end{eqnarray}
where $\epsilon_{FRW}$ and $\eta_{FRW}$ are the standard slow-roll
parameters in absence of brane corrections. The influence of brane
corrections becomes specially important when $V/\lambda_B>>1$. In
this case, we have,
\begin{equation}
\epsilon\simeq \epsilon_{FRW}(V/\lambda_B)^{-1},~~\eta \simeq 2
\eta_{FRW}(V\lambda_B)^{-1}
\end{equation}
which tells us that slow-roll ($\epsilon,\eta <<1$)is possible when
$V/\lambda_B>>1$ even if the potential is steep
($\epsilon_{FRW},\eta_{FRW}>1$). As the field rolls down its
potential, the high energy brane correction to Friedmann equation
disappears giving rise to the natural exit from inflation.

It is possible to choose potentials suitable to quintessential
inflation and fine tune the model parameters such that the model
respects nucleosynthesis constraints and leads to observed late time
cosmic acceleration\cite{samioc1,samioc2}. However, the problem
occurs  on the other side. Recent measurements of CMB anisotropies
place fairly strong constraints on inflationary models. The tensor
to scalar ratio of perturbations turns out to be lager than its
observed value in case of steep brane world inflation. Clearly, the
brane world unification of inflation and dark energy is ruled out by
observation.

\section{Modified theories of gravity and late time acceleration}
The second approach to late time acceleration is related to the
modification of
 left hand side of Einstein equations or the geometry of space time.
It is perfectly legitimate to investigate the possibility of late
time acceleration due to modification of Einstein-Hilbert action
 In the past
 few years, several schemes of large scale modifications
 have been actively investigated. Some of these modifications
are inspired by fundamental theories of high energy physics where as
the others are based upon phenomenological considerations.
 In what follows, we shall briefly describe the modified theories of
gravity and their relevance to cosmology.
\subsection{String curvature corrections}
It is interesting to investigate the string curvature corrections to
Einstein gravity amongst which the Gauss-Bonnet correction enjoys
special
status\cite{NOS,KM06,TS,CTS,Neupane,Cal,Sami06,Annalen,Sanyal}.
These models, however, suffer from several problems. Most of these
models do not include tracker like solution and those which do are
heavily constrained by the thermal history of universe. For
instance, the Gauss-Bonnet gravity with dynamical dilaton might
cause transition from matter scaling regime to late time
acceleration allowing to alleviate the fine tuning and coincidence
problems. Let us consider the low energy effective action,
\begin{eqnarray}
S&=&\int{\rm{d}}^{4}x\sqrt{-g}\Big[\frac{1}{16 \pi G}R
-(1/2)g^{\mu\nu}\partial_{\mu}\phi\:
\partial_{\nu}\phi-\nonumber \\
&-&V(\phi)-f(\phi)R_{GB}^2 \Big]+ S_{m} \label{Saction}
\end{eqnarray}
 where $ R_{GB}^2$ is the Gauss-Bonnet term,
\begin{eqnarray}
&& R_{GB}^2  \equiv R^2-4R_{\mu\nu}R^{\mu\nu}+
R_{\alpha\beta\mu\nu}R^{\alpha\beta\mu\nu}
\end{eqnarray}
The dilaton potential $V(\phi)$ and its coupling  to curvature
$f(\phi)$ are given by,
\begin{eqnarray}
&&V(\phi) \sim e^{(\alpha \phi)},~~f(\phi) \sim e^{-(\mu
 \phi)}
\end{eqnarray}
The cosmological dynamics of system (\ref{Saction}) in FRW
background was investigated in Ref.\cite{TS,KM06}. It was
demonstrated that scaling solution can be obtained in this case
provided that $\mu= \alpha$. In case $ \mu\ne \alpha$, the de-Sitter
solution is a late time attractor. Hence, the string curvature
corrections under consideration can give rise to late time
transition from matter scaling regime. Unfortunately, it is
difficult to reconcile this model with
nucleosynthesis\cite{TS,KM06}constraint.
\subsection{DGP model}
In DGP model, gravity behaves as four dimensional at small distances
but manifests its higher dimensional effects at large distances. The
modified Friedmann equations on the brane lead to late time
acceleration. The model has serious theoretical problems related to
ghost modes and superluminal fluctuations. The combined observations
on background dynamics and large angle anisotropies reveal that the
model performs much worse than $\Lambda CDM$\cite{roy}. However,
generalized versions of DGP can be ghost free and can give rise to
transient acceleration as well as a phantom phase\cite{shtanov}.
\subsection{f(R) theories of gravity}
On purely phenomenological grounds, one could seek a modification of
Einstein gravity by replacing the Ricci scalar in Einstein-Hilbert
action by $f(R)$. The action of  $f(R)$ gravity is given
by\cite{FRB},
\begin{equation}
\label{fraction}
S = \int\left[\frac{f(R)}{16\pi G} + \mathcal{L}_m \right] \sqrt{-g}
\quad d^4 x,
\end{equation}
 The modified Einstein equations which follow from (\ref{fraction}) have the form,
\begin{equation}\label{eq:freqn}
 f'R_{\mu\nu}-\nabla_{\mu}\nabla_{\nu}{f'}+\left(\Box f' - \frac{1}{2}f\right)g_{\mu\nu}
=8\pi G T_{\mu\nu}.
\end{equation}
which are of  fourth order for a non-linear function f(R). Here
prime denotes the derivatives with respect to $R$. The Ricci scalar
in FRW background is given by
\begin{equation}
R=12 H^2+6 \dot{H} \label{Ricci}
\end{equation}
which tells us that the modified Eq.(\ref{eq:freqn}) contains
de-Sitter space time as a vacuum solution provided that $
f(4\Lambda)=2\Lambda f'(4\Lambda)$. The $f(R)$ theories of gravity
may indeed provide an alternative to dark energy. To see this, let
us write the evolution equations which follow from (\ref{eq:freqn})
in a convenient form
\begin{eqnarray}
\label{EEqR1}
&&H^2=\frac{8\pi G}{3f'}\rho_R \\
&&\frac{\ddot{a}}{a}=-\frac{4\pi G}{f'}\left(\rho_R+3P_R\right)
\label{EEqR2}
\end{eqnarray}
where $\rho_R$ and $P_R$ are energy density and pressure contributed
by curvature modification
\begin{eqnarray}
\rho_R=\frac{R f'-f}{2}-3H\dot{R}f''\\
P_R=2H\dot{R}f''+\ddot{R}f''+ \frac{1}{2}(f-f' R)+ f'''\dot{R}^2
\end{eqnarray}
$\rho_R$ and $P_R$ identically vanish in case of Einstein-Hilbert
action, $f(R)=R$ as it should be. As an example of $f(R)$ model let
us consider, $f(R)=R-\alpha_n/R^n$, where $\alpha_n$ is constant for
given $n$. In case of a power law solution $a(t)\sim t^n$, the
effective equation of state parameter can be computed as
\begin{equation}
w_R=-1-\frac{2}{3}\frac{\dot{H}}{H^2}=-1+\frac{2(n+2)}{3(2n+1)(n+1)}
\end{equation}
Choosing a particular value of $n$, we can produce a desired
equation of state parameter for dark energy.

The functional form of $f(R)$ should satisfy certain requirements
for the consistency of the modified theory of gravity. The stability
of $f(R)$ theory would be ensured provided that, $f'(R)>$0 and
$f''(R)>0$ which means that graviton is not ghost and scalar degree
is not tachyon. We can understand the stability conditions
heuristically without entering into their detailed investigations.
From evolution equations (\ref{EEqR1}) $\&$ (\ref{EEqR1}), we see
that the effective gravitational constant $G_{eff}=G/f'$ which
should be positive or $f'>0$ in order to avoid the pathological
situation. As for the second condition, V. Faraoni has given an
interesting interpretation\cite{faraoni}: let us consider the
opposite case when $f''<0$ which means that $G'_{eff}=-f''G/f'^2>0$.
This implies that gravitational constant increases for increasing
value of $R$ making the gravity stronger. In view of Einstein
equations, it leads to yet larger value of curvature and so on which
ultimately leads to a catastrophic situation. Thus we need $f''$ to
be positive to avoid the catastrophe.

Let us note that $f(R)$ gravity theories apart from a spin two
object necessarily contain a scalar degree of freedom. Taking trace
of Eq.(\ref{eq:freqn}) gives the evolution equation for the scalar
degree of freedom,
\begin{equation}\label{eq:frtrace}
 \Box f' = \frac{1}{3} \left ( 2f' - f' R \right ) + \frac{8\pi G}{3} T.
\end{equation}
It should be noticed that Eq.(\ref{eq:frtrace}) reduces to an
algebraic relation in case of Einstein gravity; in general $f'$ has
dynamics. It is convenient to define scalar function $\phi$ as,
\begin{equation}
 \phi \equiv f' - 1,
\end{equation}
which is expressed through Ricci scalar once $f(R)$  is specified.
 We can write the trace equation (Eq.(\ref{eq:frtrace}))
 in the terms of $V$ and $T$ as
\begin{equation}
 \Box \phi = \frac{dV}{d\phi} +\frac{8\pi G}{3} T.
\end{equation}
which is a Klein-Gordon equation in presence of a deriving term.
Thus $\phi$ is indeed a scalar degree of freedom which controls the
curvature of space time.

The effective potential can be evaluated using the following
relation
\begin{equation}\label{Effpot}
 \frac{dV}{dR} = \frac{dV}{d\phi}\frac{d\phi}{dR}= \frac{1}{3}
\left ( 2 f - f' R \right )
 f''.
\end{equation}
Models which satisfy the stability conditions belong to two
categories: (1) Either they are not distinguishable from $\Lambda
CDM$ or are not viable cosmologically. (ii) Models with disappearing
cosmological constant: In these models, $f(R)\to 0$ for $R\to 0$ and
they give rise to cosmological constant in regions of high density
and differ from the latter otherwise. In principal, these models can
be distinguished from cosmological constant. Models belonging to the
second category were proposed by Hu-Sawicki and Starobinsky
\cite{HS,star}(see also Ref.\cite{Appl} on the similar theme). The
functional form of $f(R)$ in Starobinsky parametrization is given
by,
\begin{equation}  \label{func}
f(R) = R + \lambda R_0 \left[\left(1+\frac{R^2}{R_0^2}\right)^{-n}
-1\right].
\end{equation}
 Here $n$ and  $\lambda$ are positive. And $R_0$ is of the
order of presently observed cosmological constant, $\Lambda = 8\pi G
\rho_{vac}$. The model satisfies the stability conditions quoted
above.

 In the Starobinsky model, the scalar field $\phi$, in the absence of matter, is given by
\begin{equation}\label{eq:phi}
 \phi(R) = -\frac{2n \lambda R}{R_0(1+\frac{R^2}{R_0^2})^{n+1}}.
\end{equation}
Notice that $R \to \infty$ for $\phi \to 0$. For a viable late time
cosmology, the field should be evolving near the minimum of the
effective potential. The finite time singularity inherent in the
class of models under consideration severely constrains dynamics of
the field.

\begin{figure}[htp]
\includegraphics[width=71mm]{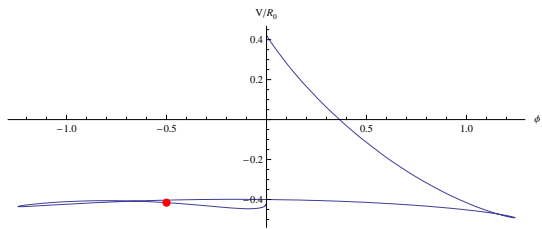}
\caption{Plot of effective potential for $n = 2$ and $\lambda =
1.2$. The red spot marks the initial condition for
evolution.}\label{POT}
\end{figure}

\subsection*{The curvature singularity and fine tuning of parameters}
The effective potential has minimum which depends upon $n$ and
$\lambda$. For generic values of the parameters, the minimum of the
potential is close to $\phi =0$(see Fig.\ref{POT}) corresponding to
infinitely large curvature. Thus while the field is evolving towards
minimum, it can easily oscillate to a singular
point\cite{Apl2,frolov}. However, depending upon the values of
parameters, we can choose a finite range of initial conditions for
which scalar field $\phi$ can evolve to the minimum of the potential
without hitting the singularity. We find that the range of initial
conditions allowed for the evolution of $\phi$ to the minimum
without hitting singularity shrinks as the numerical values of
parameters $n$ and $\lambda$ increase. In the presence of matter,
the minimum of the effective potential moves towards the origin. In
case of the compact objects such as neutron stars, the minimum is
extremely near the origin and the singularity problem becomes really
acute\cite{frolov,maeda1}.
\subsection*{Avoiding singularity with higher curvature corrections}
We know that in case of large curvature, the quantum effects become
important leading to higher curvature corrections. Keeping this in
mind, let us consider the modification of Starobinsky's
model\cite{TH,maeda2,Odsing},
\begin{equation}
 f(R) = R + \frac{\alpha}{R_0} R^2 +R_0 \lambda \left \lbrack -1 + \frac{1}{(1+\frac{R^2}{R_0^2})^n} \right \rbrack,
 \label{R2}
\end{equation}
then $\phi$ becomes
\begin{equation}\label{eq:phi2}
 \phi(R) = \frac{R}{R_0}\left[2\alpha  -\frac{2n \lambda }{(1+\frac{R^2}{R_0^2})^{n+1}}\right].
\end{equation}
In case  $|R|$ is large, the first term which comes from $\alpha
R^2$ dominates. In this case, the curvature singularity, $R = \pm
\infty$
 corresponds to $\phi = \pm \infty$. Hence, in this modification, the
minimum of the effective potential is separated from the curvature
singularity by the infinite distance in the $\phi,V(\phi)$ plane.
  Though the introduction of $R^2$ term
formally allows to avoid the singularity but can not alleviate the
fine tuning problem as the minimum of the effective potential should
be near the in generic cases. As for the compact objects, Langlois
and Babichev\cite{Lango}(see, Ref.\cite{NOOD} also on the similar
theme) have argued that neutron stars can be rescued from
singularity if a realistic equation of state for these objects is
used though the numerical simulation is yet challenging for
densities of the order of nuclear matter density. The problem
deserves further investigation.

In scenarios of large scale modification of gravity, one should
worry about the local gravity constraints. The $f(R)$ theories are
related to the class of scalar tensor theories corresponding the
Brans-Dicke parameter $\omega=0$ or the PPN parameter
$\gamma=(1+\omega)/(2+\omega)=1/2$ unlike GR where $\gamma=1$
consistent with observation ($|\gamma-1|\lesssim 2.3 \times
10^{-5}$). This conclusion can be escaped by invoking the so called
chameleon mechanism\cite{TBrax}. In case, the scalar degree of
freedom is coupled to matter, the effective mass of the field
depends upon the matter density which can allow to avoid the
conflict with solar physics constraints. However, the problem of
singularity in these models is genuine and should be addressed.
\section{Summary}
We have given a pedagogical exposition of physics of late time
cosmic acceleration. Most of the part of the review should be
accessible to a graduate student. The discussion of Newtonian
cosmology is comprehensive and reviews the efforts to put the
formalism of Newtonian cosmology on rigorous foundations in its
domain of validity. Heuristic discussion on the introduction of
cosmological constant and pressure corrections in evolution
equations is included. The underlying idea leading to late time
cosmic acceleration is explained without the use of general theory
of relativity. The basic features of cosmological dynamics in
presence of cosmological constant is presented in a simple and
elegant fashion making it accessible to non-experts. The review also
gives the glimpses of relativistic cosmology, contains important
notes on the dynamics of dark energy and discusses underlying
features of cosmological dynamics of a variety of scalar fields
including quintessence, rolling tachyon and phantom. Special
emphasis is put on the cosmic viability of these models; the
cosmological relevance of scaling solutions is briefly explained.
The review ends with a discussion on modified theories of gravity as
possible alternatives to dark energy. The treatment is simple but
conveys the successes and problems of cosmology in the frame work of
modified theories of gravity. Basic features of $f(R)$ cosmology are
explained avoiding the cumbersome mathematical expressions. The
latest developments of $f(R)$ theories with disappearing
cosmological constant are highlighted. The problem of singularities
in these
models and their possible resolution are discussed. \\
I hope the review would be helpful to beginners and will also be of
interest to experts.
\section{Acknowledgements}
I am indebted to T. Padmanabhan for giving me an opportunity to
write this review for a special issue of Current Science. I am
thankful to D. Jain, V. Sahni and I. Thongkool  for taking pain in
going through the manuscript and making suggestions for its
improvement. I also thank S. A. Abbas, N. Dadhich, S. Jhingan, R. P.
Kirshner, K. Raza and A. A. Sen for useful comments and discussion.

\end{document}